\documentclass[aps,prd,12pt,nofootinbib]{revtex4}

\usepackage{epsfig}
\usepackage{graphicx}
\usepackage{adjustbox}
\usepackage[font=small,skip=0pt,justification=raggedright]{caption}
\usepackage{amsmath}
\usepackage{amsthm}
\usepackage{amssymb}
\usepackage{mathrsfs}
\usepackage{verbatim}
\usepackage{makecell}
\usepackage{booktabs}

\usepackage[normalem]{ulem}
\usepackage{xcolor}

\newcounter{fig}   

\newcommand{\Tr}{{\rm Tr}}

\newcommand{\bea}{\begin{eqnarray}}
\newcommand{\eea}{\end{eqnarray}}
\newcommand{\be}{\begin{equation}}
\newcommand{\ee}{\end{equation}}

\newcommand{\re}[1]{(\ref{#1})}

\newcommand{\eqn}{\begin{eqnarray}}
\newcommand{\eqnx}{\end{eqnarray}}

\date{\today}

\begin{document}
\title{Stress stability criterion of the isospinning $\mathbb{C}P^2$ solitons}
%
%
\author{Sergei Antsipovich}
\affiliation{
Department of Theoretical Physics and Astrophysics,
Belarusian State University, Minsk 220004, Belarus
}
\author{Ya.~Shnir}
\affiliation{BLTP, JINR, Dubna 141980, Moscow Region, Russia
}

\begin{abstract}
We study the  energy-momentum tensor of stationary rotating topological solitons in a (2 + 1)-dimensional $\mathbb{C}P^2$  nonlinear sigma model
with a stabilizing potential term. We evaluate the distributions of the corresponding shear forces and pressure and study the
stability criteria for these solutions. Our results suggest that these solitons become classically unstable for some range of values of the parameters of the system. 
\end{abstract}
\maketitle

\section{Introduction}
In various nonlinear field theories, spatially localized regular solutions known as solitons arise. Many such models have attracted considerable attention in recent few decades in a wide variety of physical contexts.
Solitons are relevant to numerous areas of physics;
they naturally arise in condensed matter physics, classical
and quantum field theories, cosmology, biology, nuclear
physics, and other disciplines. In many situations, the existence of solitons is determined by the topological properties of the system; see, e.g., \cite{Manton:2004tk,Shnir2018}.

One of the most well-known examples are topological solitons in the (2+1)-
dimensional non-linear $\mathbb{C}P^N$ sigma model, they were considered for the
first time in the papers \cite{Golo:1978de,DAdda:1978vbw,Din:1980jg,Zakrzewski}. The simplest
$\mathbb{C}P^1$ model is equivalent to the $O(3)$ nonlinear sigma model \cite{Leese:1989gi,Sutcliffe:1991aua},
soliton solutions of this conformally invariant self-dual theory were constructed in \cite{Polyakov:1975yp}.
A few modifications of the $\mathbb{C}P^N$ sigma model were proposed to obtain solitons which
do not saturate the topological bound. In particular, the theory can be modified by inclusion
into Lagrangian some additional terms, which are higher order in derivatives, and an appropriate potential term.
A basic example of such a deformation is the planar Skyrme model
\cite{Bogolubskaya:1989ha,Bogolyubskaya:1989fz,Piette:1994jt,Piette:1994ug}.

Another mechanism of stabilization of solitons, which allows to circumvent Derrick's theorem \cite{Derrick:1964ww},
is related with internal rotations of the configuration. Such time-dependent topological solitons were constructed
both in the $\mathbb{C}P^1$ model with symmetry breaking potential \cite{Abraham:1991ki,Ward:2003un}, and in the
$\mathbb{C}P^2$ model \cite{Amari:2024pnw,Antsipovich:2025liy}. Certainly, there
is a similarity between  such internally rotating topological solitons \cite{Radu:2008pp}
and Q-balls, which are time-dependent
lumps of a complex scalar field with a stationary oscillating phase \cite{Rosen,Friedberg:1976me,Coleman:1985ki}.

It should be noted that the harmonic time dependence of a topological soliton can lead to a
change in its shape and breaking the symmetries of the static configuration.
In general, isorotating solitons cannot be approximated as a rigid body
\cite{Halavanau:2013vsa,Battye:2013tka,Battye:2013xf,Harland:2013uk,Battye:2014qva}. Further, deformations of
topological solitons may result in instability of the configurations with respect to decay into
constituents \cite{Halavanau:2013vsa,Battye:2013tka}.
It should be noted that topological arguments alone are not sufficient to guarantee the stability of solitons:
they can contract or expand if they are not balanced by other physical mechanisms.
Depending on the properties of the potential and the values of the model parameters, stable, metastable and unstable solutions may exist.

It was noted that the problem of soliton stability can be analyzed by studying the matrix elements of the energy-momentum tensor and the associated spatial distributions of forces acting within the configuration 
\cite{Polyakov:2002yz,Polyakov:2018zvc,Mai:2012yc,Mai:2012cx,Loiko:2022noq,Panteleeva:2023aiz,Farakos:2025byy,Mikhaliuk:2026tdt}.
This approach supplements so-called Vakhitov-Kolokolov criteria of stability of solitons
with respect to linearized perturbations \cite{Vakhitov:1973lcn}, it originates from related study of
form factors of the energy-momentum tensor of hadrons \cite{Polyakov:2002yz,Polyakov:2018zvc} and evaluation
of the corresponding \textit{D-term}, the quantity which describes the 
 internal mechanical structure of localized configurations, specifically the distribution of pressure and shear
forces. It should be noted that
this method can be used for both topological \cite{Panteleeva:2023aiz,Farakos:2025byy}
and non-topological \cite{Mai:2012yc,Mai:2012cx,Loiko:2022noq,Mikhaliuk:2026tdt} solitons. The condition of stability of the localized  configuration with respect to internal deformations is the negativity of the D-term.

In this paper we extend this approach to the case of stationary rotating topological solitons in a
(2+1)-dimensional $\mathbb{C}P^2$ model with a stabilizing potential term \cite{Amari:2024pnw,Antsipovich:2025liy}.
We study the energy-momentum tensor and discuss corresponding distributions of the
pressure and shear forces, acting in the interior of the isorotating $\mathbb{C}P^2$ soliton.

This paper is organized as follows. In Sec. II, we introduce the model and
summarize the properties of isospinning topolocally nontrivial $\mathbb{C}P^2$
solitons. Sec. III provides analysis of the stress tensor and the
problem of the distribution of the shear forces and pressure
acting on the configurations. Numerical results
are presented in Sec. IV, where we discuss the stability
condition of the $\mathbb{C}P^2$ solitons which follow from the
conservation of the energy-momentum tensor of the system.
Conclusions and remarks are formulated in Sec. V.

\section{The model}
We consider a $2+1$
dimensional nonlinear  sigma model with the target space $\mathbb{C}P^{2} \backsimeq
SU(3)/[SU(2)\times U(1)]$ \cite{Golo:1978de,DAdda:1978vbw,Din:1980jg,Zakrzewski}.
The  Lagrangian of the $\mathbb{C}P^2$ model is
\be \label{lag}
{\cal L} =\frac14 \Tr (\partial_\mu \mathbf{n}\, \partial^\mu \mathbf{n}) - V(\mathbf{n})
\, ,
\ee
where the field $\mathbf{n}= n^a\lambda_a$, $(a=1,2\dots 8)$ and $\lambda_a$ are the usual $SU(3)$ Gell-Mann matrices
subject to normalization $\Tr (\lambda_a \lambda_b) = 2\delta_{ab}$.
The components of the field $n^a$ satisfy the constraints:
\be
n^a n^a =\frac{4}{3}\, , \qquad n^a=\frac{3}{2}\, d_{abc} n^b n^c\,
\label{const}
\ee
where the totally symmetric third rank tensor  $d_{abc}=\frac14 \Tr(\lambda_a\{\lambda_b,\lambda_c\})$ yield the second
$SU(3)$ Casimir operator, $d_{abc}n^a n^b n^c = 8/9$.

The first term in \re{lag} is  the usual Lagrangian of the
non-linear sigma-model. It is invariant with respect to the global transformation of the
field $\mathbf{n} \to U^\dagger  \mathbf{n} U$, where $U\in ~SU(3)$.
The vacuum symmetry $SU(2)\times U(1)$ is explicitly broken by
the potential $V(\mathbf{n})$, which is necessary to stabilize the configuration. As in previous works
\cite{Amari:2024pnw,Antsipovich:2025liy},
we consider a potential which breaks this symmetry to the Cartan subgroup $H=U(1)\times U(1)$.

Topologically, the field is defined as a map
$\mathbf{n}: S^2 \to \mathbb{C}P^{2}$ characterized by an element of the second homotopy group
$Q_{\rm top} \in \pi_2(\mathbb{C}P^{2}) =
\mathbb{Z}$. Explicitly, the topological charge is an integer defined by the integral
\be
Q_{\rm top}=\frac{1}{8\pi}\int d^2x \varepsilon^{ij} f_{abc} n^a \partial_i n^b \partial_j n^c
\label{topcharge}
\ee
Here $f_{abc}= \frac{i}{4}\Tr(\lambda_a[\lambda_b,\lambda_c])$ are the $SU(3)$ structure constants.

Since, taking into account two constrains \re{const}, the model \re{lag} possesses six
physical degrees of freedom, it is more convenient to introduce a complex  3-component vector $Z$
which satisfies $Z^\dagger Z =1$ and defines the homogeneous coordinates on $\mathbb{C}P^{2}$. Here the dagger indicates
Hermitian conjugation. The components of this vector are related to the field $\mathbf{n}$ as
$$
n^a = Z^\dagger \lambda_a Z \, .
$$
In terms of $Z$ the Lagrangian \re{lag} becomes
\be
{\cal L}= 2 (D_\mu Z)^\dagger (D^\mu Z) - V(Z)
\label{lagZ}
\ee
while the topological charge \re{topcharge} is given by
\be
Q_{\rm top}=-\frac{i}{2\pi}\int d^2 x \varepsilon^{ij} (D_i Z)^\dagger (D_j Z)=
\int d^2 x \varepsilon^{ij} \partial_i A_j
\label{topZ}
\ee
Here, the covariant derivative is defined as
$D_\mu Z=\partial_\mu Z - i A_\mu Z  $ with the induced connection
$A_\mu=-i Z^\dagger \partial_\mu Z$. Hence,
\be
(D_\mu Z)^\dagger (D^\mu Z) = \partial_\mu Z^\dagger \,
\partial^\mu Z - (\partial_\mu Z^\dagger  Z)
(Z^\dagger \partial^\mu  Z)\, .
\ee

The corresponding stress-energy tensor is
\be
T_{\mu\nu} = 2 \left\{ (D_\mu Z)^\dagger (D_\nu Z) + (D_\nu Z)^\dagger (D_\mu Z) \right\}- {\cal L} g_{\mu\nu}\, .
\label{energy-stress}
\ee
Thus, the energy (mass) of the configuration is given by
\be
M=\int d^2 x ~{\cal{E}} = \int d^2 x\left[2\left\{
(D_0 Z)^\dagger (D_0 Z) + (D_i Z)^\dagger (D_i Z)\right\} + V(Z)\right]
\label{T00}
\ee
where ${\cal{E}}= T_{00}$ is the energy density
and the angular momentum of the spinning configuration reads
\be
J=\int d^2 x~ \mathcal{J}  =2 \int d^2 x \left[(D_0 Z)^\dagger (D_\theta Z)
+ (D_\theta Z)^\dagger (D_0 Z) \right] \, ,
\label{J}
\ee
where $\mathcal{J} = T_{0\theta}$ and the variables $(r,\theta)$ are the usual polar coordinates on the plane $\mathbb{R}^2$.

As in the previous discussion of isorotating $\mathbb{C}P^{2}$ solitons \cite{Amari:2024pnw}, we consider
$U(1)\times U(1)$ rotationally-invariant Ansatz for the fundamental field
\be Z=\left(\cos F(r),~\sin F(r) \cos G(r)e^{i \varphi},~\sin F(r) \sin G(r)e^{i
\psi} \right)^\tau\, ,
\label{Z}
\ee
where $\varphi=\omega_1 t + k_1 \theta$ and $\psi=\omega_2 t + k_2 \theta$ with angular
frequencies $\omega_{1,2}$ and integer winding numbers $k_{1,2}$. The two corresponding
$U(1)$ generators with the left/right  action on $Z$ are
\be
t_\varphi = ~{\rm diag}~\{0,1,0\},\quad
t_\psi = ~{\rm diag}~\{0,0,1\}\, . \label{u1gens}
\ee

Clearly, under transformation $\theta \to \theta + \delta
\theta$, the fundamental field $Z$ varies as
\be
\begin{split}
Z &\to e^{i k_1 t_\varphi \delta \theta }e^{i k_2 t_\psi \delta \theta } Z =
e^{\frac{i}{3} (k_1+k_2) \delta \theta }
e^{-\frac{i k_1}{2} \lambda_3 \delta \theta }
e^{\frac{i}{2 \sqrt3} (k_1-2k_2)\lambda_8 \delta \theta }\,Z\\
&=e^{-\frac{i k_1}{2} \lambda_3 \delta \theta }
e^{\frac{i}{2 \sqrt3} (k_1-2k_2)\lambda_8 \delta \theta }\,Z ~~~({\rm mod~global~}U(1))\, .
\end{split}
\ee
while the components of the octet $\mathbf{n}$ transform as
\be
\begin{split}
(n^1+in^2) \mapsto (n^1+in^2)e^{i\omega_1 t}, ~~ &(n^4+in^5) \mapsto (n^4+in^5)e^{i\omega_2 t},
~~(n^6+in^7) \mapsto (n^6+in^7)e^{i(\omega_1-\omega_2) t},\\
& n^3 \mapsto n^3, ~~\quad n^8 \mapsto n^8 \, .
\end{split}
\label{rotcomp}
\ee

The corresponding conserved Noether currents are
\be
\begin{split}
j_\mu^{(\varphi)} &= -n^1\partial_\mu n^2 + n^2 \partial_\mu n^1 +n^6\partial_\mu n^7 - n^7 \partial_\mu n^6\, ,
\\
j_\mu^{(\psi)} &=-n^4\partial_\mu n^5 + n^5 \partial_\mu n^4 -n^6\partial_\mu n^7 + n^7 \partial_\mu n^6\, .
\end{split}
\label{Noether-j}
\ee
They yield two Noether charges,
\be
\begin{split}
Q_\varphi &= \int d^2 x j_0^{(\varphi)} =
2\pi \int rdr \left\{\omega_1 \sin^2(2F)\cos^2 G
+ (\omega_1-\omega_2)\sin^4 F \sin^2(2G) \right\}  \, , \\
Q_\psi &= \int d^2 x j_0^{(\psi)} = 2\pi \int rdr \left\{
\omega_2 \sin^2(2F)\sin^2 G - (\omega_1-\omega_2)\sin^4 F \sin^2(2G)
 \right\} \, ,
\end{split}
\label{2Q}
\ee
respectively.

Note that the spacial rotations of a planar $\mathbb{C}P^2$ configuration  are generated by a single
Killing vector field $\partial /\partial \theta$, the corresponding
angular momentum is given by
\begin{align}
J&= \int d^2 x T_{0\theta}
\notag\\
&=2 \pi \int dr \left\{ \sin^2(2 F)[\omega_1 k_1 \cos^2 G + \omega_2 k_2 \sin^2 G ]
+(k_1-k_2)(\omega_1-\omega_2) \sin^4 F \sin^2(2G)\right\}\, \notag\\
&=k_1 Q_\varphi + k_2 Q_\psi\, ,
\label{moment}
\end{align}
where the two Noether charges \re{2Q} are defined above.
We can write the associated total energy \eqref{T00} of the configuration \re{Z} as
\begin{align}
M=&\int d^2x T_{00}
\notag\\
=& 2\pi\int rdr \left[2 \left\{ (F^\prime)^2 + (G^\prime)^2 \sin^2 F \right\} + \frac12 \sin^4 F \sin^2(2G)\left\{ \frac{(k_1-k_2)^2}{r^2} + (\omega_1-\omega_2)^2 \right\} \right.
\notag \\
&\left. \qquad\qquad\qquad\qquad\quad+
\frac12 \sin^2(2 F)\left\{ \cos^2 G \left( \frac{k_1^2}{r^2} + \omega_1^2\right) +\sin^2 G \left( \frac{k_2^2}{r^2} + \omega_2^2\right) \right\}
+ V\right] \, ,
\label{TotEng}
\end{align}
where the prime stands for the derivative with respect to the radial coordinate $r$.
Moreover, substituting the Ansatz \re{Z} into Eq.~\re{topZ},
one finds that the topological charge can be rewritten in the form
\be
Q_{\rm top}=\int d^2 x \varepsilon^{ij} \partial_i A_j = 
\left[\sin^2 F(r)\left\{k_1 \cos^2 G(r) +k_2 \sin^2 G(r)\right\}
\right]^{r\to \infty}_{r=0} \ ,
\label{topcharge2}
\ee
Hence, its value is determined by the boundary conditions imposed on the
profile functions at the origin and on the spatial infinity.

Our choice of the symmetry breaking potential term $V(Z)$ is motivated by this structure of the
frequency-dependent terms in the expression for the energy.
In particular, the term $(\partial_0 n^a)^2$  must be vanishing at least at one of vacua of the potential.
Using the Ansatz \eqref{Z}, we obtain
\be
\begin{split}
(\partial_0 n^a)^2 &= \frac19\biggl[ \omega_1^2\left(2-3 n^3 + \sqrt3 n^8 \right)(2+3n^3 + \sqrt3 n^8)\\
&+2\omega_2^2(1-\sqrt3 n^8)\left(2+3n^3 +\sqrt3 n^8\right)
+2(\omega_1-\omega_2)^2 (1-\sqrt3 n^8)(2-3n^3+\sqrt3 n^8)
\biggr]     \, .
\end{split}
\label{iso_density}
\ee
Here we follow the choice of \cite{Amari:2024pnw}, where
a corresponding double vacuum potential is defined as
\be
V=\mu^2 \arctan (5 W(n^3, n^8)) \, .
\label{Pot-CP2}
\ee
where
\be
W(n^3, n^8)=\left((2+ 3 n^3 + \sqrt 3 n^8)(2- 3 n^3 + \sqrt 3 n^8)\right)(n^8)^2 + (1-\sqrt 3 n^8) \, ,
\label{pot2}
\ee
It has the two minima $(n^3,n^8) = (\pm 1, 1/\sqrt{3})$. The corresponding effective potential is
\begin{equation}
    V_{\rm eff}(n^3,n^8)= (\partial_0 n^a)^2 -  V \ .
    \label{Veff}
\end{equation}

\begin{figure}[!h]
	\centering
	\adjincludegraphics[trim={{.0\width} {.0\width} {.0\width} {.0\width}},clip, width=0.34\textwidth]{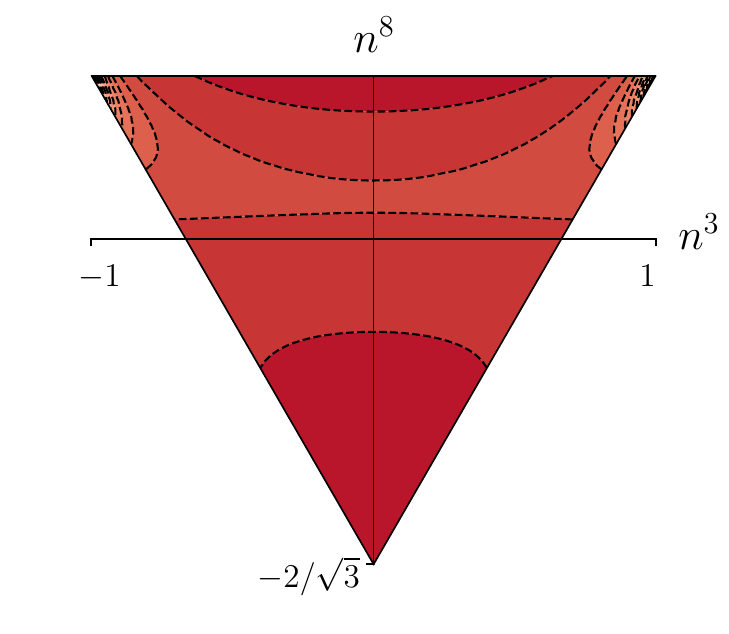}%
	\adjincludegraphics[trim={{.0\width} {.0\width} {.4\width} {.0\width}},clip, width=0.137\textwidth]{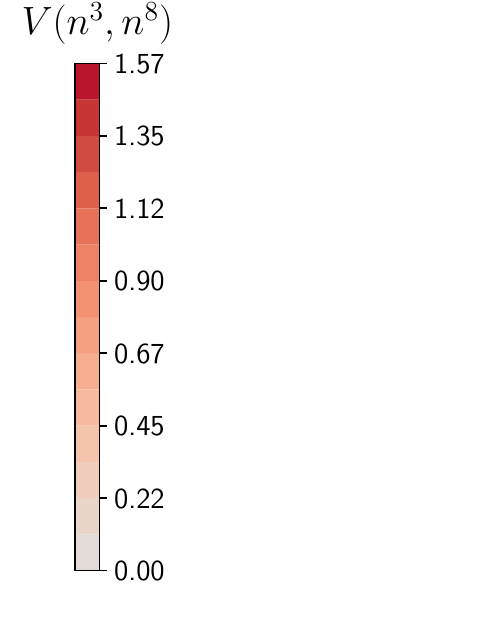}%
    \adjincludegraphics[trim={{.0\width} {.0\width} {.0\width} {.0\width}},clip, width=0.34\textwidth]{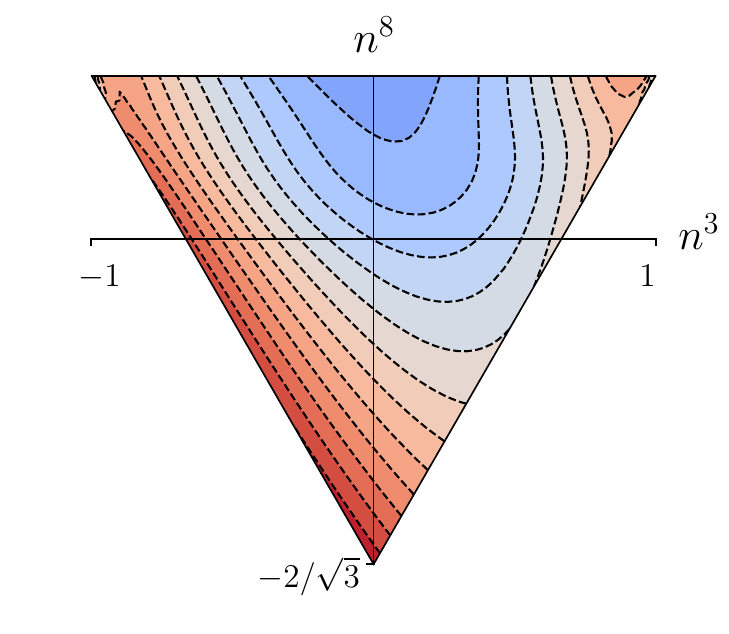}%
	\adjincludegraphics[trim={{.0\width} {.0\width} {.4\width} {.0\width}},clip, width=0.137\textwidth]{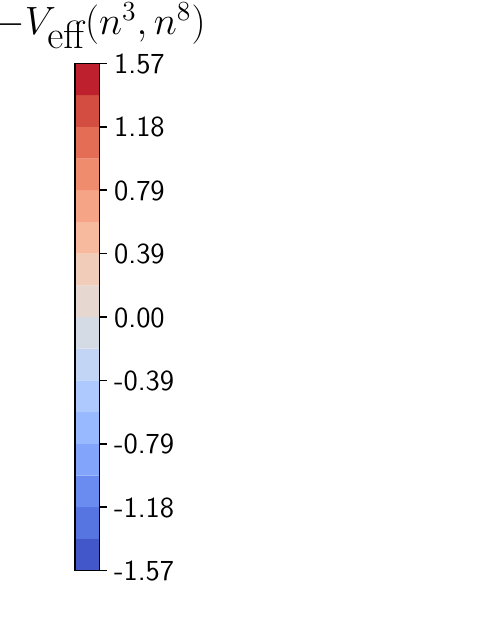}%
	\caption{Contour plots of the potential  \re{Pot-CP2}  (left panel) and the effective potential \re{Veff}  (right panel) are displayed for $\omega_1=2.2$, $\omega_2 =1.7$ and $\mu=1$.}
	\label{pot_and_effpot}
\end{figure}


Notably, the effective potential becomes non-positive for some range of values of the frequencies
$(\omega_1\, , \omega_2)$. This feature is a necessary condition to circumvent restrictions of the
Derrick's theorem securing the existence of stable solitons in flat space,
see e.g. \cite{Verbin:2007fa,Kleihaus:2013tba,Ferreira:2025xey}.

The two profile functions of the configuration are subject to the
following boundary conditions \cite{Amari:2024pnw}
\begin{equation}
    n^3(0)=1, ~n^8(0)=1/\sqrt 3, ~n^3(\infty)=-1, ~n^8(\infty)=1/\sqrt 3 \ ,
    \label{BC1}
\end{equation}
or
\be
F(0)=0, ~~~G(0)=\pi/2, \quad
F(\infty)=\pi/2, ~~~G(\infty)=0 \, .
\label{BC1_FG}
\ee
Substitution of this boundary conditions into the expression for the topological charge \re{topcharge2}
gives
\begin{equation}
    Q_{\rm top}=k_1 \ .
\end{equation}
Hence, for such a choice of the boundary conditions, the second integer $k_2$, which parameterize the
Ansatz \eqref{Z},  remains a free non-topological parameter\footnote{This distinction between the winding numbers is not fundamental whatsoever. By choosing a potential with suitable symmetries and/or by changing the Ansatz, one may set $Q_{\textup{top}}$ to be defined by the second integer, $k_{2}$, see \cite{Antsipovich:2025liy}.}. Note that the configuration with $k_1=k_2$ can be constructed for modified boundary conditions 
\be
F(0)=0, ~~~\partial_r G(0)=0, \quad
F(\infty)=\pi/2, ~~~G(\infty)=0 \, .
\label{BC1_FG_k1}
\ee
For all solutions satisfying the boundary conditions \re{BC1_FG}, $k_1 \ge k_2$ \cite{Amari:2024pnw}. 

The asymptotic expansion of the Euler-Lagrange equations at $r\to0$ yields

\begin{equation}
     F(r)=\varepsilon f(r), \qquad G(r)=\frac{\pi}{2}-\varepsilon g(r)
\end{equation}

produces a set of two Cauchy-Euler equations in $O\left(\varepsilon^{2}\right)$ and $O\left(\varepsilon^{4}\right)$ orders of approximation, respectively:

\begin{equation}
\begin{split}
f^{\prime \prime} &+ \frac{f^\prime}{r} - k_2^2\,\frac{f}{r^{2}}=0,\\
g^{\prime \prime} &+ \left(1+2|k_{2}|\right)\frac{g^\prime}{r} -
\left(k_{1}^{2}-k_{2}^{2} \right)\frac{g}{r^{2}} =0 \, ,
\end{split}
\end{equation}

with the solutions 

\begin{equation}
     f(r)=C_{\!f}\,r^{|k_{2}|}, \qquad g(r)=C_{\!g}\,r^{(|k_{1}|-|k_{2}|)}\;.
\end{equation}

A similar result may be obtained from examining the Lagrangian \eqref{lag} in vacuum near the spatial origin. One may see, that since both the potential and the isorotational term in the Lagrangian become zero in $F(0),G(0)$ \eqref{BC1_FG}, the model reduces into $\mathcal{L}=2 \left ( D^{j}Z \right )^{\dagger }D_{j}Z$ \cite{Golo:1978de, DAdda:1978vbw}, with the only field configurations satisfying the ansatz \eqref{Z} being

\begin{equation*}
     Z_{\textup{BPS}}=\frac{\left (1,az^{k_{1}},bz^{k_{2}}  \right )^{T}}{\sqrt{1+\left |az^{k_{1}} \right |^2+\left |bz^{k_{2}} \right |^2}}\;.
\end{equation*}

Assuming $k_{1}\geqslant k_{2}$, one may get

\begin{equation}
\begin{split}
F(r)&=\arccos\left(\frac{1}{\sqrt{1+a^{2}r^{2k_{1}}+b^{2}r^{2k_{2}}}}\right)\approx C_{\!f} r^{k_{2}}+ \dots,\\ G(r)&=\arccos\left(\frac{ar^{k_{1}}}{\sqrt{a^{2}r^{2k_{1}}+b^{2}r^{2k_{2}}}}\right)\approx \frac{\pi}{2}-C_{\!g} r^{(k_{1}-k_{2})}+ \dots
\end{split}
\end{equation}

Similarly, the asymptotic expansion of the fields around the vacuum on the spacial infinity 
$$
F(r)= \pi/2 - \varepsilon f(r), \qquad G(r) = \varepsilon g(r) 
$$
yields two linearized equations for the fluctuations of the fields 
\be
\begin{split}
f^{\prime \prime} &+ \frac{f^\prime}{r} - (30 \mu^2 - \omega_1^2)f=0,\\
g^{\prime \prime} &+ \frac{g^\prime}{r} -
\left(\frac{15}{2} \mu^2 - (\omega_1-\omega_2)^2 \right)g =0 \, .
\end{split}
\ee
Hence, the mass of excitations of the components $F$ and $G$ is different, 
$$
m_f=\sqrt{30 \mu^2 - \omega_1^2},\qquad m_g=\sqrt{\frac{15}{2} \mu^2 - (\omega_1 - \omega_2)^2}
$$
Therefore, spatially localized solitonic configurations may exist if the frequencies are restricted from above as 
\be
\left\{%
\begin{array}{ll}
    |\omega_1| \le \sqrt{30}\, \mu \\
    |\omega_1-\omega_2| \le \sqrt{\frac{15}{2}} \, \mu \\
\end{array}%
\right.
\ee
Note that these conditions do not depend on the values of the integers $k_1,k_2$.

By analogy with the usual Q-balls, there is also a lower limit to the range of allowed values of the  frequencies. It follows from the condition that the effective potential 
\re{pot_and_effpot} must possess at least one
nodal point to support stable solitons. Numerical calculations lead to an estimate of the range of acceptable frequency values, as displayed in Fig.\ref{figs_Ew1w2_k1_1_k2_1}. Note that the angular frequency $\omega_1$ can also take negative values.

Further, the necessary linear stability
condition  of a classical isorotating $\mathbb{C}P^{2}$ soliton
can be formulated as a generalized Vakhitov-Kolokolov criterion \cite{Vakhitov:1973lcn},
the boundary between stable and unstable regimes
is characterized by the vanishing of the corresponding Jacobian determinant \cite{Acus:2012ff}
\be
\left\| \frac{\partial(Q_\phi,Q_\psi)}{\partial (\omega_1,\omega_2)} \right\| =0 \, .
\label{VK}
\ee

\section{Mechanical properties of $\mathbb{C}P^{2}$ solitons }
It was suggested by Maxim Polyakov \cite{Polyakov:2002yz,Polyakov:2018zvc} that
soliton-like configurations may be considered by analogy with droplets of an elastic media. Explicitly, the spacial components of the stress tensor can associated
with the distribution of the pressure anisotropy (shear forces) $s(r)$ and the elastic
pressure $p(r)$ inside the system,
\be
\label{stressPartEMTmetric}
        T_{jk}=-p(r) g_{jk}+s(r)\left (\frac{x_{j}x_{k}}{r^2}+\frac{g_{jk}}{2} \right )\, ,
\end{equation}
where $g_{\mu\nu}=\textup{diag}\left ( 1,-1,-r^2 \right )$ is  the flat space metric tensor in (2+1) dimensions.
For the $\mathbb{C}P^{2}$ solitons the energy-stress tensor is defined as \re{energy-stress}, thus
\begin{equation}\label{system_2}
\begin{cases}
T_{rr}= 2\left ( \left ( D_{0}Z \right )^{\dagger }D_{0}Z
+\left ( D_{r}Z \right )^{\dagger }D_{r}Z-\frac{1}{r^2}\left ( D_{\theta}Z \right )^{\dagger }D_{\theta}Z \right )
-V  \\
T_{\theta\theta}= 2r^2\left ( \left ( D_{0}Z \right )^{\dagger }D_{0}Z-\left ( D_{r}Z \right )^{\dagger }D_{r}Z
+\frac{1}{r^2}\left ( D_{\theta}Z \right )^{\dagger }D_{\theta}Z \right )-r^{2} V\, .
\end{cases}
\end{equation}
Thus,
\begin{equation}
\label{system_3}
\begin{cases}
 p(r)=2\left ( D_{0}Z \right )^{\dagger }D_{0}Z-V \\
 s(r)=4\left ( D_{r}Z \right )^{\dagger }D_{r}Z-\frac{4}{r^{2}}
 \left ( D_{\theta}Z \right )^{\dagger }D_{\theta}Z.
\end{cases}
\end{equation}
Substituting the explicit form of the Ansatz \eqref{Z}  into these equations,
we obtain the following expressions for the pressure and shear force distributions
\begin{equation}
\label{pressre_FG}
\begin{split}
        p(r)=\frac{1}{2}\left(\omega _1-\omega _2\right)^2 \sin ^4(F) \sin ^2(2 G)+\frac{1}{2}\sin ^2(2 F)
        \left(\omega _2^2 \sin ^2(G)+\omega _1^2 \cos ^2(G)\right)-V
\end{split}
\end{equation}
and
\begin{equation}\label{shear_FG}
\begin{split}
   s(r)=4(F')^2+4\sin ^2(F)\; (G')^2-\frac{\left(k_1-k_2\right)^2 \sin ^4(F) \sin ^2(2 G)}{r^2} \\
   -\frac{\sin ^2(2 F) \left(k_2^2 \sin ^2(G)+k_1^2 \cos ^2(G)\right)}{r^2}
\end{split}
\end{equation}
Here the potential $V$ is defined by \re{Pot-CP2},\re{pot2}.

Thus, combining definitions \eqref{T00}, \eqref{J}, and \eqref{stressPartEMTmetric}, we arrive at the following explicit form of the energy-momentum tensor \eqref{energy-stress} of the stationary spherically symmetric  configurations:
\begin{equation}\label{explicit_EMT}
     T_{\mu\nu}=\begin{pmatrix}
\mathcal{E} & 0 & \mathcal{J} \\
0 & p\!+\!\frac{s}{2} & 0 \\
\mathcal{J} & 0 & r^2(p\!-\!\frac{s}{2}) \\
\end{pmatrix}
\end{equation}
where $\mathcal{E}=T_{00}$ and $\mathcal{J}=T_{0\theta}$.

The conservation of this (2+1)-dimensional stress-energy tensor
implies that the pressure and shear force distributions satisfy the following \emph{equilibrium} equation
\be
d(r)=p^\prime(r)+\frac{1}{2}s^\prime(r)+\frac{1}{r}s(r)=0\, .
\ee
It implies that
\be
p^\prime(r)=-\frac{1}{2r^2}\frac{\mathrm{d}}{\mathrm{d}r}\left [r^{2} s(r) \right ]
\label{stability-d}
\ee

Another restriction is the von Laue  condition \cite{Laue:1911lrk,Polyakov:2018zvc,Pinto:2025plg},
related with the repulsive and attractive forces balance in the interior of a soliton:
\be
\int d^2x\, p(r) = 0
\ee
Consequently, the pressure function $p(r)$ must possess at least one zero.
This is a necessary (though not sufficient)
condition of stability, it can be also reformulated as a virial relation connecting
integrals of components of the energy-momentum tensor.

Imposing the finite upper integration limit $R$ and integrating the von Laue condition
by parts, we obtain
\be
\int_{0}^{R} r dr\, p(r)= \left[ \frac{r^{2}}{2}p(r) \right]_{0}^{R}
-\int_{0}^{R}dr\, \frac{r^{2}}{2}p^\prime(r)
\ee
Using the relation \re{stability-d}, we arrive at
\be
\int_{0}^{R}rdr\, p(r)r =\left [ \frac{r^{2}}{2}\left ( p(r) +\frac{1}{2}s(r)
\right ) \right ]_{0}^{R}
\label{int-Laue}
\ee
The expression in brackets in \re{int-Laue} corresponds to
the distribution of the radial component of the net force
acting on an infinitesimal area element $dA\, \boldsymbol{\hat{e}}_{r}$ at a distance $r$
\cite{Polyakov:2018zvc,Perevalova:2016dln}, $dF_{j}=T_{jk}\, dA_{k}$ and the normal force must be directed outward
\be
C(r)=\frac{dF_{r}}{dA_{r}}=p(r) +\frac{1}{2}s(r) \ge 0
\label{C-criterion}
\ee

Further, by analogy with investigation of
internal structure of hadrons, one can introduce one more
mechanical characteristic, so-called \emph{D-term} (or the \emph{Druck-term}) which is related to
the distribution of the internal forces inside the configuration \cite{Polyakov:2002yz,Choudhary:2022den}.
It can be expressed in terms of the pressure and shear force distributions as
\be
D(0)=8\pi M \int_{0}^{\infty}rdr \, r^{2} p(r)=-2\pi M \int_{0}^{\infty}rdr\, r^{2} s(r) \, ,
\label{D-term}
\ee
where $M$ is the mass  of the stationary configuration \re{T00}.  
Since the pressure and the shear force distributions are restricted by relation \re{C-criterion},
for any stable system the D-term must be negative \cite{Polyakov:2002yz,Polyakov:2018zvc,Perevalova:2016dln}.

Further, it is quite convenient to consider a dimensionless quantity \cite{Mai:2012cx}
\begin{equation}
    \tilde{D}=\frac{ D(0)}{ M^{2}\langle r^{2}\rangle },
\end{equation}
where  $\langle r^{2}\rangle$ is the mean square radius of the soliton, defined as
\be
\langle r^{2}\rangle=\frac{\int d^2x\, r^{2}T_{00}(r)}{\int d^2x\, T_{00}(r)}.
\ee

To complete our discussion of mechanical properties of the $\mathbb{C}P^{2}$ solitons we consider the energy conditions
that are usually applied in the general  relativity. They follow from the assumption that the energy density should
be positive everywhere in space. Explicitly, the following energy conditions must hold:
\begin{itemize}
    \item Weak energy condition (WEC) $ T_{\mu\nu}X^{\mu}X_{\nu}\geqslant 0$ for every timelike vector field
    $X_\mu$;
    \item Null energy condition (NEC) $ T_{\mu\nu}Y^{\mu}Y_{\nu}\geqslant 0$ for every null (lightlike) vector field
    $Y_\mu$;
    \item Strong energy condition (SEC) $\left ( T_{\mu\nu}-Tg_{\mu\nu} \right )X^{\mu}X^{\nu}\geqslant 0$ for every timelike vector field
    $X_\mu$;
    \item  Dominant energy condition (DEC) $j_{\mu}j^{\mu}\geqslant 0$, where $X_\mu$ is a timelike vector field and $j_{\mu}=-T_{\mu\nu}X^{\nu}$ is the energy flux vector field.
\end{itemize}
Considering isospinning $\mathbb{C}P^{2}$ soliton as a (2+1)-dimensional droplet of a rotating viscous fluid with
energy density ${\cal E}$ \re{T00}, angular momentum density $\mathcal{J}$ \re{J} shear forces distribution $s$ and pressure $p$,
we can select, for example,
a timelike vector $X_\mu=(1,0,0)$.
Then the WEC becomes ${\cal E} \ge 0$, i.e. for an observer,
moving along $X_\mu$, the energy density cannot be negative. Similarly, for a null vector we can pick
$Y_\mu=(1,1,0)$, it yields the NEC ${\cal{E}}+p+\frac{1}{2}s \ge 0$. Further, the SEC represent a statement that the
force of interaction is locally attractive, etc. All energy  relations of the $\mathbb{C}P^{2}$ solitons
are summarized in Table \ref{ec-table}. Using explicit expressions for all the terms appearing in the inequalities above, one can show that they all hold, with the exception of the SEC condition for timelike vectors.

\begin{table}
 \renewcommand{\arraystretch}{1.2}
\begin{tabular}{ c @{\qquad} c @{\qquad} c }
\hline
\hline
\textbf{} &  \makecell{Timelike,  $X=\left(1,0,0\right)$} & \makecell{Lightlike,  $Y=\left(1,1,0\right)$} \\
\colrule
\textbf{WEC}  & ${\cal{E}}\geqslant 0 $ &  ${\cal{E}}+p+\frac{s}{2} \geqslant 0$
 \\
 \colrule
    \textbf{DEC}  & ${\cal{E}}^2-\frac{\mathcal{J}^{2}}{r^{2}}\geqslant 0$ & ~~ ${\cal{E}}^2-\frac{\mathcal{J}^2}{r^2}-(p+\frac{s}{2})^2\geqslant 0$
 \\
 \colrule
    \textbf{SEC}  & $p \geqslant 0$ &  ${\cal{E}}+p+\frac{s}{2} \geqslant 0$
 \\
\hline
\hline
\end{tabular}
\vspace{7pt}
\caption{Energy conditions for the $\mathbb{C}P^{2}$ solitons}
  \label{ec-table}
\end{table}

\section{Numerical results}

\subsection{Numerical method}

\begin{figure}[!htb]
	\centering
	\includegraphics[width=0.57\textwidth]{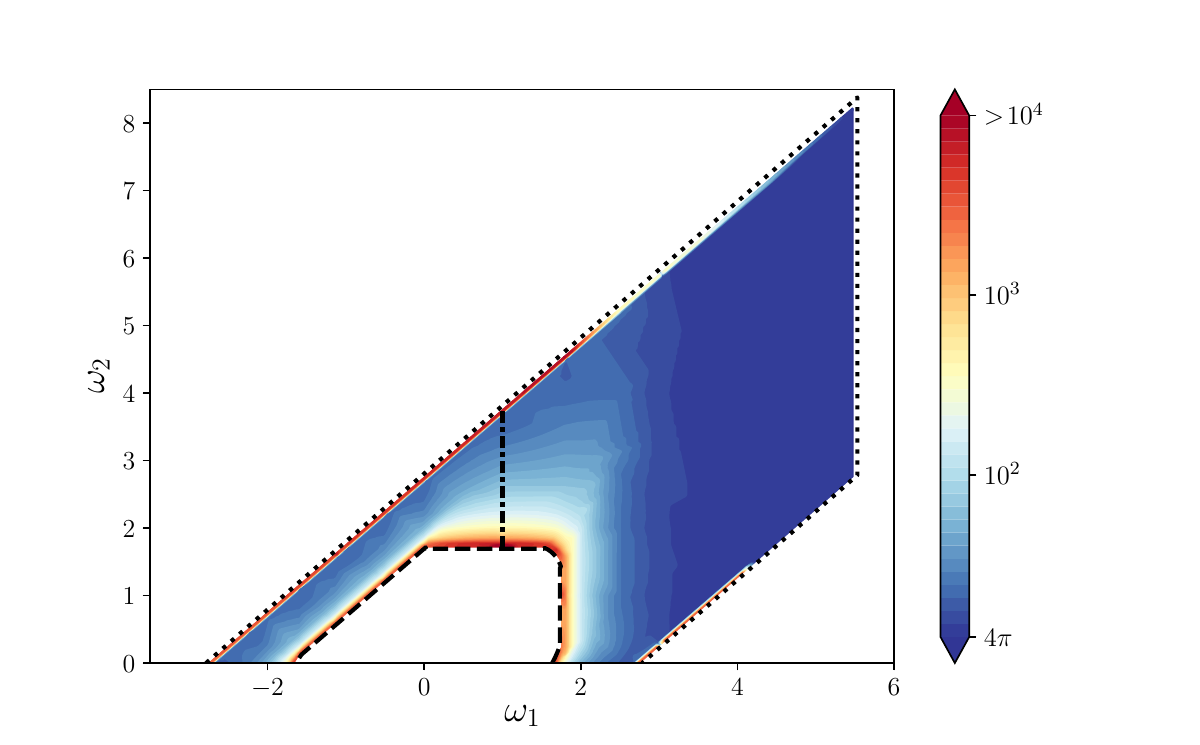}
\includegraphics[width=0.41\textwidth]{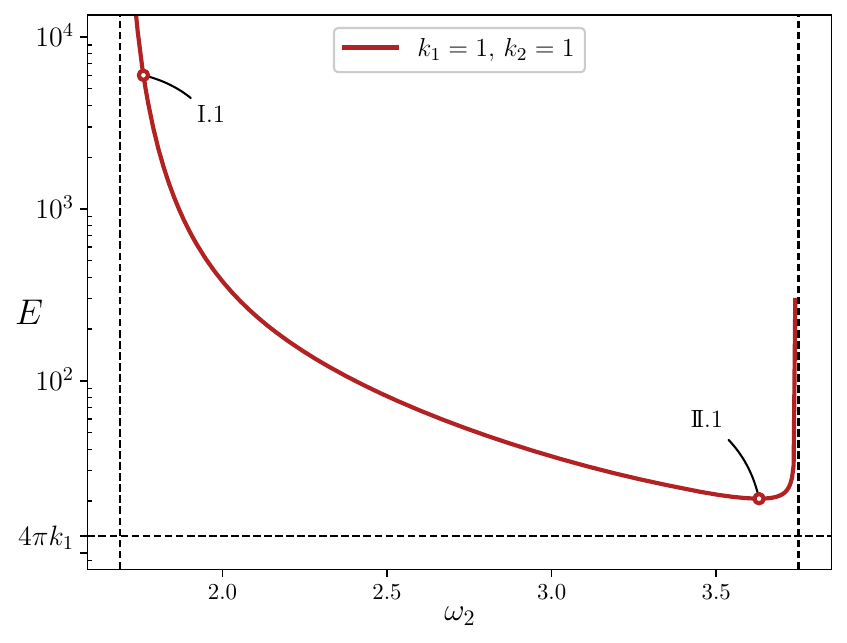}%

	\caption{Left plot: Contour map of the total energy distribution of the isospinning $\mathbb{C}P^{2}$ solitons is displayed  as function of $\omega_{1}$ and $\omega_{2}$ over a domain of acceptable frequencies for $\mu=1$ and $k_{1}=k_{2}=1.$ The vertical dashed-dotted line corresponds to the scan shown in the right plot. }\label{figs_Ew1w2_k1_1_k2_1}
\end{figure}

To find stationary points of the energy functional \re{TotEng} we solved the corresponding system of variational equations with the boundary conditions \re{BC1_FG}. As a consistency check, we verify that our algorithm agrees with the virial relation between the potential and
quadratic in derivatives terms in the 
effective  energy functional \re{TotEng} within 1.0$\%$ accuracy. In our numerical scheme we implement the sixth-order
finite-difference method. The system of equations is discretized on an equidistant grid in radial coordinate $x=\frac{r/r_0}{1+r/r_0}\in [0,1]$, where $r_0$ is a real
scaling constant, which typically is  taking values from 0.5 to 15. Estimated numerical
errors are of order of $10^{-5}$. 

The input parameters are the mass parameter
$\mu$, the winding numbers $k_1,k_2$ and the angular frequencies $\omega_1,\omega_2$, restricted by the  domain of existence. Since the parameter space of the system is quite large, we will restrict our analysis to the case $\mu=1$ and consider, as  two particular examples, configurations of topological degrees one and four, i.e. with
$k_1=1$ and $k_1=4$ and allowed corresponding values of $k_2$. 

\subsection{Mechanical properties of $Q_{\textup{top}}=1$ $\mathbb{C}P^2$ soliton}

\begin{figure}[!htb]
	\centering
	\includegraphics[width=0.45\textwidth]{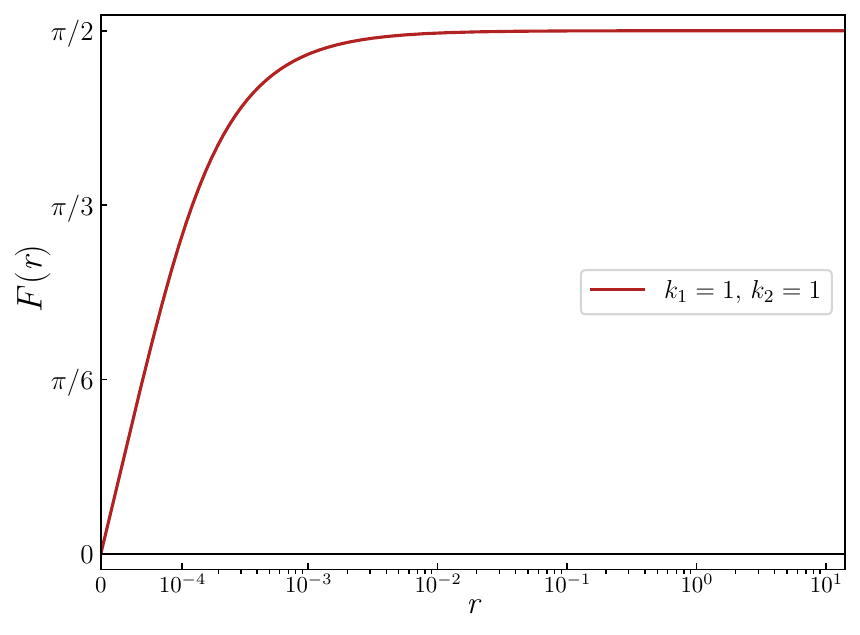}%
	\includegraphics[width=0.45\textwidth]{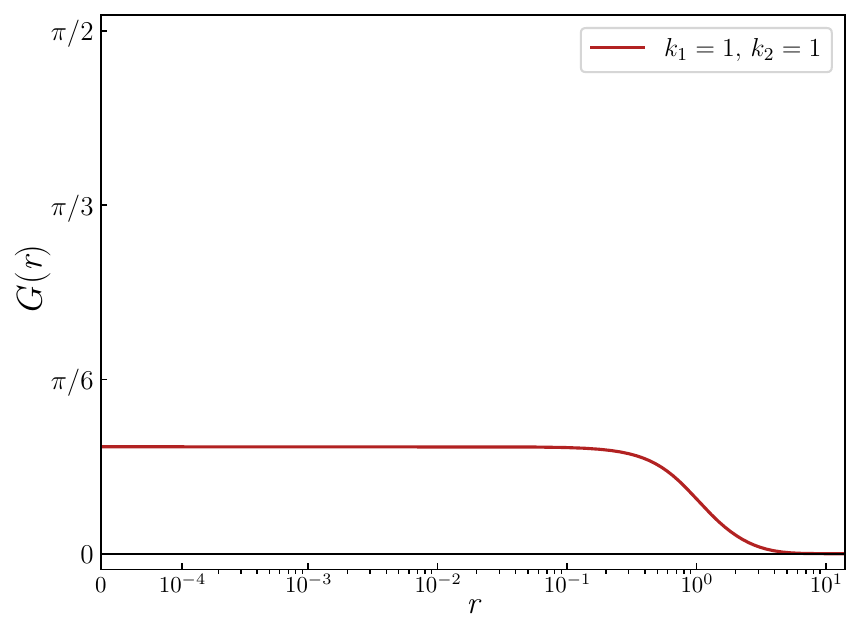}
	\caption{Radial profile functions of  the $k_{1}=1$ soliton at $\omega_1 =1$ and $\omega_2=3.625 $, labeled as $\textup{II}.1$ on
the Fig.~\ref{figs_Ew1w2_k1_1_k2_1}, right plot  (thick-wall limit) .}\label{FG_1n_thick-wall_profiles}
\end{figure}

\begin{figure}[!htb]
	\centering
\includegraphics[width=0.45\textwidth]{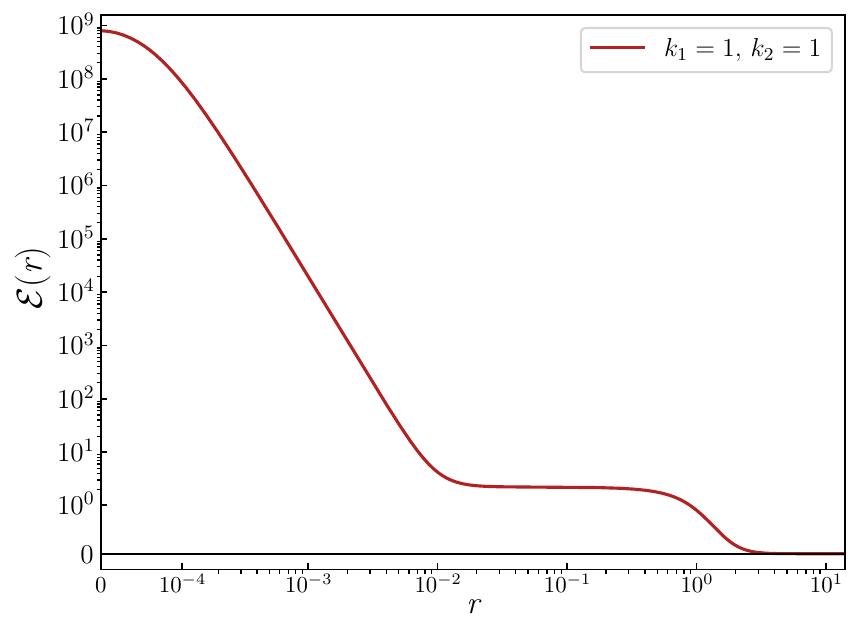}%
\includegraphics[width=0.45\textwidth]{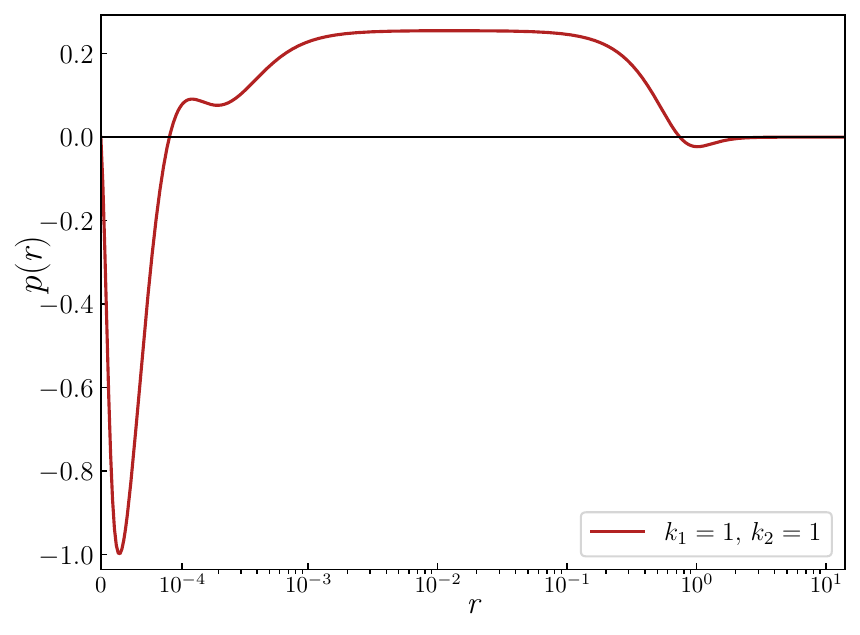}%

\includegraphics[width=0.45\textwidth]{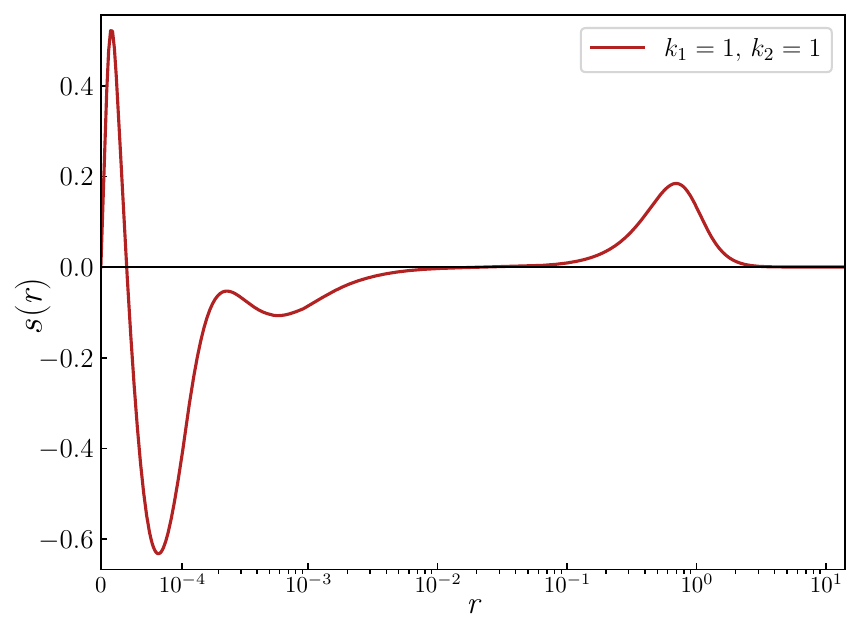}%
\includegraphics[width=0.45\textwidth]{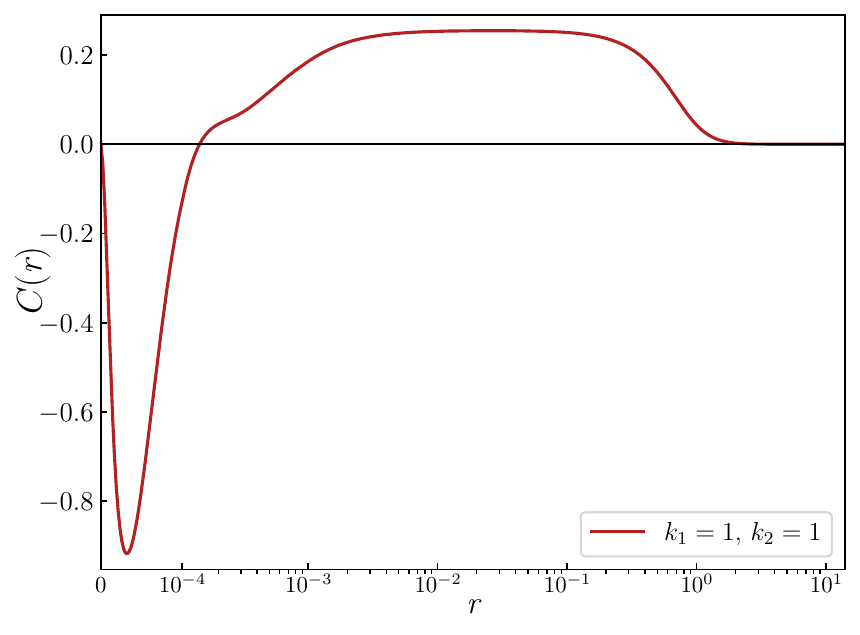}

	\caption{Radial profiles of the energy density distribution $\mathcal{E}(r)$ (upper left plot), the pressure function $p(r)$ (upper right plot), and shear force $s(r)$ (bottom left plot),  and the radial force component $C(r)$ (bottom right plot)  of the particular $k_{1}=k_2=1$ solution at $\omega_1 =1, \omega_2=3.625 $, labeled as $\textup{II}.1$ on
the Fig.~\ref{figs_Ew1w2_k1_1_k2_1}, right plot. 
}\label{ps_1n_thick-wall_profiles}
\end{figure}

Let us briefly review the basic properties of the non-trivial topological soliton solutions of the model \re{lag}  discussed
previously in \cite{Amari:2024pnw,Antsipovich:2025liy}. In the sector of topological degree one there is only one solution with winding numbers $k_1=k_2=1$ which satisfies the boundary conditions \re{BC1_FG_k1}.

\begin{figure}[!htb]
	\centering
	\includegraphics[width=0.45\textwidth]{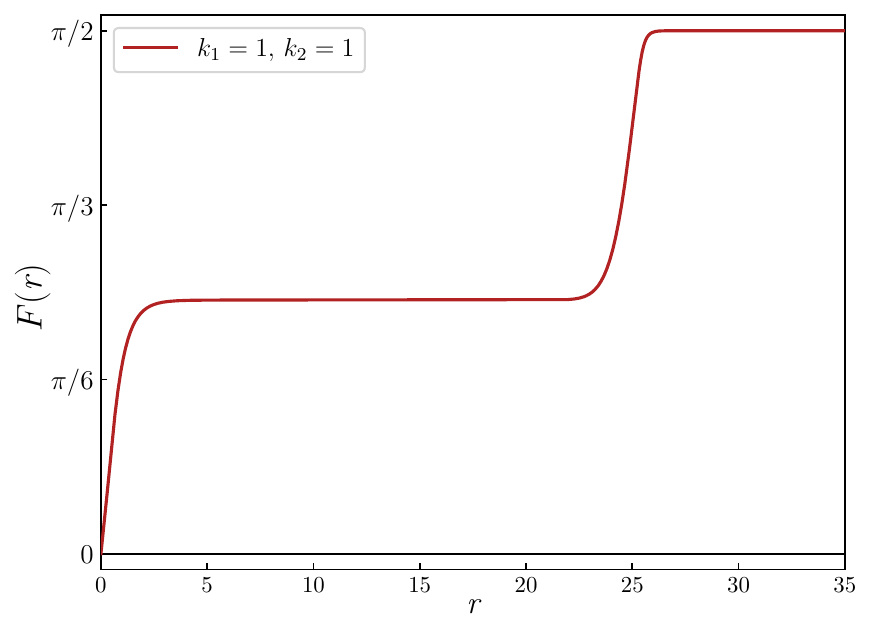}%
	\includegraphics[width=0.45\textwidth]{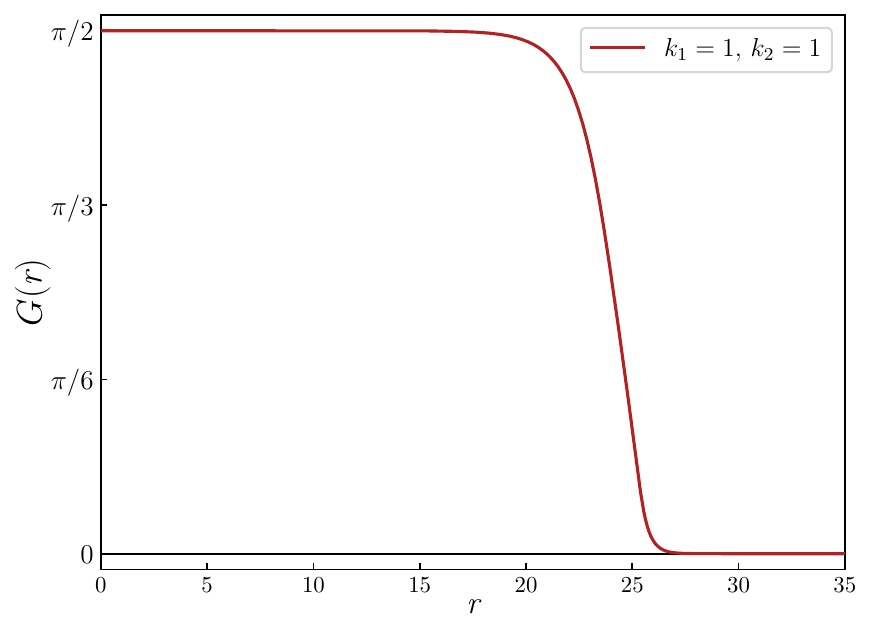}
	\caption{Radial profile functions of  the $k_{1}=k_2=1$ $\mathbb{C}P^{2}$ soliton at $\omega_1 =1$ and $\omega_2= 1.760 $, labeled as $\textup{I}.1$ in  Fig.~\ref{figs_Ew1w2_k1_1_k2_1}, right plot     
    (thin-wall limit).
    }\label{FG_1n_thin-wall_profiles}
\end{figure}

We scan the parametric space taking a particular value of
the angular frequency $\omega_1$ and varying the second frequency
$\omega_2$.
The localized isospinning soliton arises as the angular frequency 
decrease below upper critical values, and the corresponding
total energy functional develops a local minimum. Figs.~\ref{FG_1n_thin-wall_profiles},\ref{FG_1n_thick-wall_profiles}  display, as particular examples, the profile functions of  $k_1=k_2=1$ solutions for a fixed value of $\omega_1=1$ and  
$\omega_2=3.625 $ (thick-wall limit)
$\omega_2 =1.760 $ and  (thin-wall limit), respectively. Notably, the profile function $G(r)$ trivializes as $\omega_2$ approaches the upper critical value. It is also interesting to note that, in the thin-wall limit, the configuration splits into two "half-solitons", see Fig.~\ref{FG_1n_thin-wall_profiles}. 

For a given value of the frequency $\omega_1$, both the energy
and the angular momentum of the configuration increase as the second frequency
$\omega_2$ decreases,  the
size of the soliton expands rapidly as the volume energy
increases. The solitons possess
finite energy and charge for all allowed ranges of values of the
angular frequencies.

Fig.~\ref{figs_Ew1w2_k1_1_k2_1} displays a contour map of the total energy density distribution of the $k_1=k_2=1$ configurations in the range of allowed frequencies. The vertical dashed-dotted line corresponds to the scan at fixed $\omega_1=1$, shown in the right  plot.  As we see, the Vakhitov-Kolokolov inequality does not hold for the thick-wall configurations, as $\omega_2$ approaches the upper limit.

In Fig.~\ref{ps_1n_thick-wall_profiles} we exhibit the distributions of the total energy density \re{TotEng} and the functions $p(r)$ \re{pressre_FG}, $s(r)$ \re{shear_FG} and the normal force $C(r)$ \re{C-criterion} of
the particular solutions, labeled as $\textup{II}.1$ on
the Fig.~\ref{figs_Ew1w2_k1_1_k2_1}, right plot. 

We note that the radial distribution of energy density of a thick-wall soliton is positive everywhere, it is characterized by a peak at the origin and an extended plateau, as shown in Fig.\ref{ps_1n_thick-wall_profiles}, upper left plot.  The corresponding pressure distribution $p(r)$ possesses two nodes, it is negative at the very center of the soliton, then it becomes positive in the region of the extended plateau. Second node appears at the end of the plateau, as displayed in  Fig.~\ref{ps_1n_thick-wall_profiles}, upper right plot.

Consequently, the von Laue stability condition becomes weakly violated on the tail of the $\mathbb{C}P^{2}$ soliton, although the the shear forces distribution $s(r)$ and the criterion function $C(r)$ remain positive there, see Fig.~\ref{ps_1n_thick-wall_profiles}, bottom plots.
The shear force distribution $s(r)$ changes its sign twice, it has one positive region at the center of the soliton and the other positive region at the end of an extended plateau on its tail, see Fig.~\ref{ps_1n_thick-wall_profiles}, bottom left plot.

\begin{figure}[!htb]
	\centering
\includegraphics[width=0.45\textwidth]{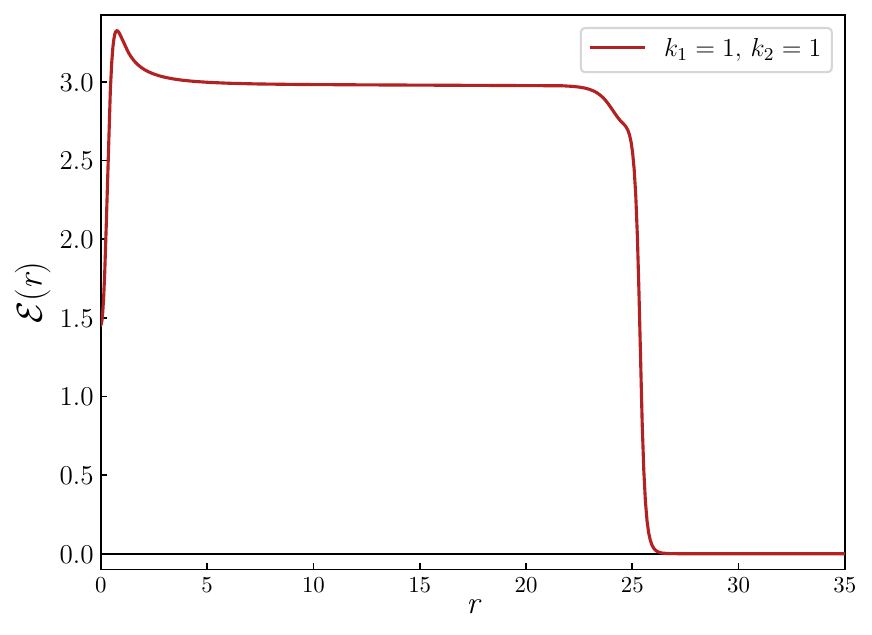}%
\includegraphics[width=0.45\textwidth]{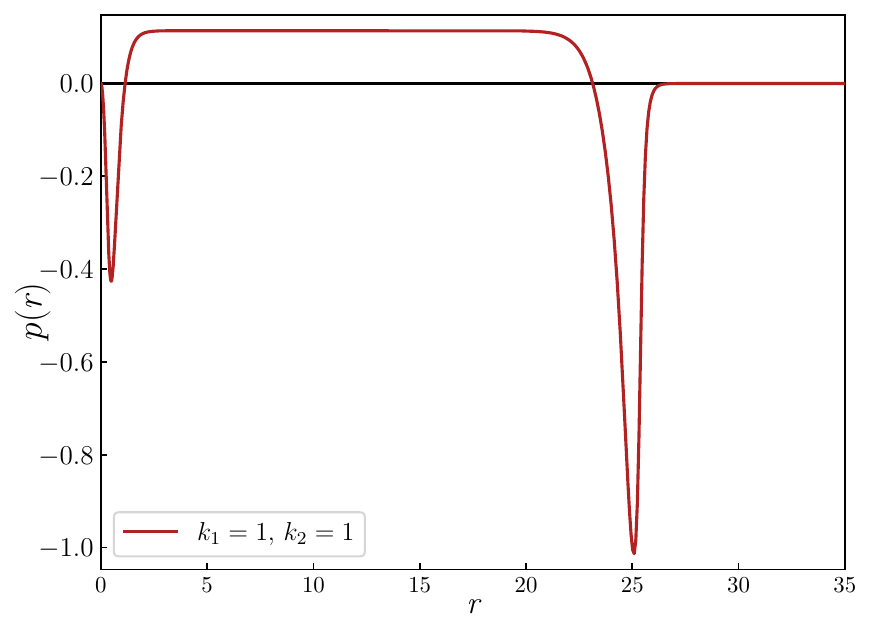}\\
	\includegraphics[width=0.45\textwidth]{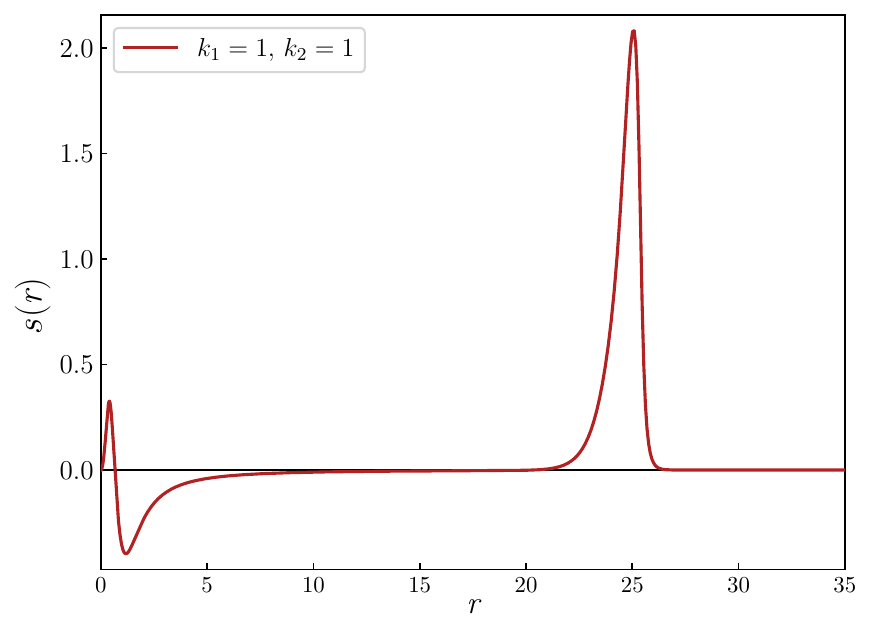}
\includegraphics[width=0.45\textwidth]{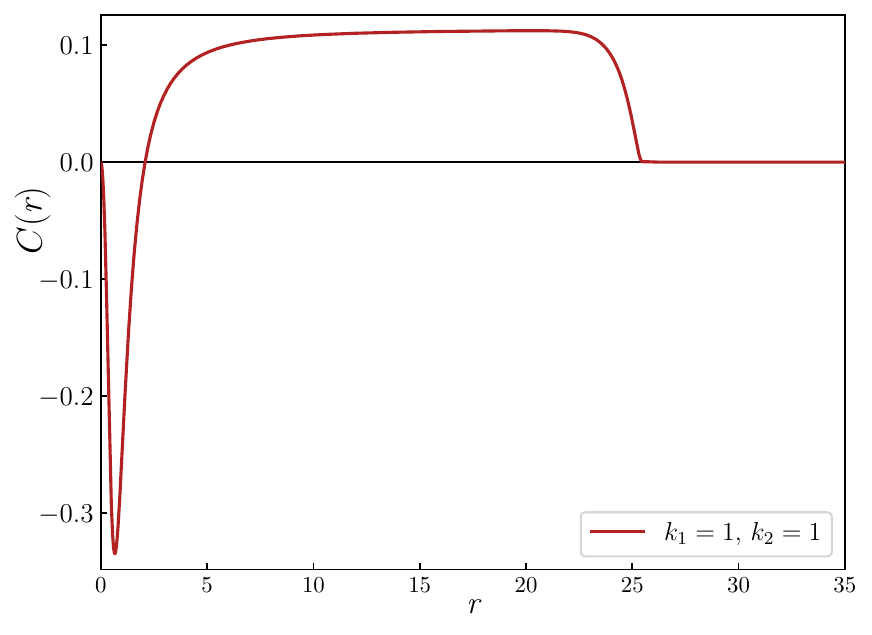}

	\caption{Radial profiles of the energy density distribution $\mathcal{E}(r)$ (upper left plot), the pressure function $p(r)$ (upper right plot), and shear force $s(r)$ (bottom left plot),  and the radial force component $C(r)$ (bottom right plot)  of the particular $k_{1}=k_2=1$ solution at $\omega_1 =1, \omega_2=1.760 $, labeled as $\textup{I}.1$ on
the Fig.~\ref{figs_Ew1w2_k1_1_k2_1}, right plot. 
}\label{fig6}
\end{figure}
\begin{figure}[!htb]
	\centering
	\includegraphics[width=0.45\textwidth]{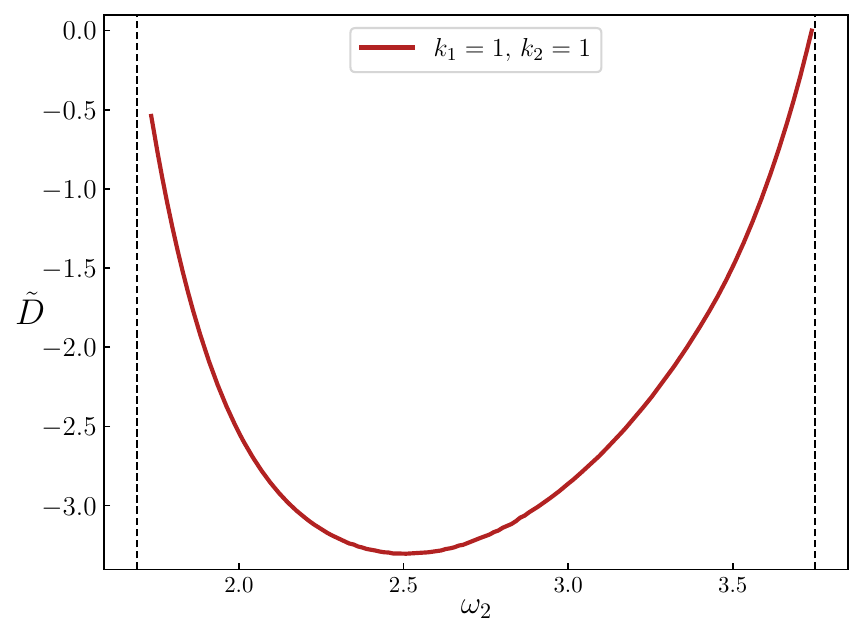}
	\caption{The normalized D-term as function of $\omega_{2}$ for the fixed $\omega_{1}=1.00$ and $k_{1}=1$.}\label{figs_E_D_k1_1_k2_n}
\end{figure}

In
Fig.~\ref{FG_1n_thin-wall_profiles} we displayed the profile functions of the $k_{1}=k_2=1$ $\mathbb{C}P^{2}$ soliton in the thin-wall limit. 
The Fig.~\ref{fig6} shows the radial distributions of the total energy density
(11) and the functions $p(r)$, $s(r)$, and $C(r)$
of the corresponding solution,
 labeled as $\textup{I}.1$ on the Fig.~\ref{figs_Ew1w2_k1_1_k2_1}, right plot.
 Clearly, the soliton of topological degree one 
in this limit splits into two “half-solitons”. Both the pressure function $p(r)$ and the shear force distribution $s(r)$ have two nodes; these functions tend to zero at the center of the configuration and at the spatial boundary. The criterion function 
$C(r)$ is negative in the interior of the soliton, it becomes positive outside the core approaching zero from above, see Fig.~\ref{fig6}, bottom right plot.  Thus, we can clearly see that both the von Laue condition and the stability criteria \re{C-criterion} are violated. The value of the D-term  for $k_1=k_2=1$ $\mathbb{C}P^{2}$ soliton at $\omega_1=1$ varies with $\omega_2$, however, it 
always remains negative, see Fig.~\ref{figs_E_D_k1_1_k2_n}.

\subsection{Mechanical properties of $Q_{\textup{top}}=4$ $\mathbb{C}P^2$ soliton}

\begin{figure}[!htb]
	\centering
	\includegraphics[width=0.45\textwidth]{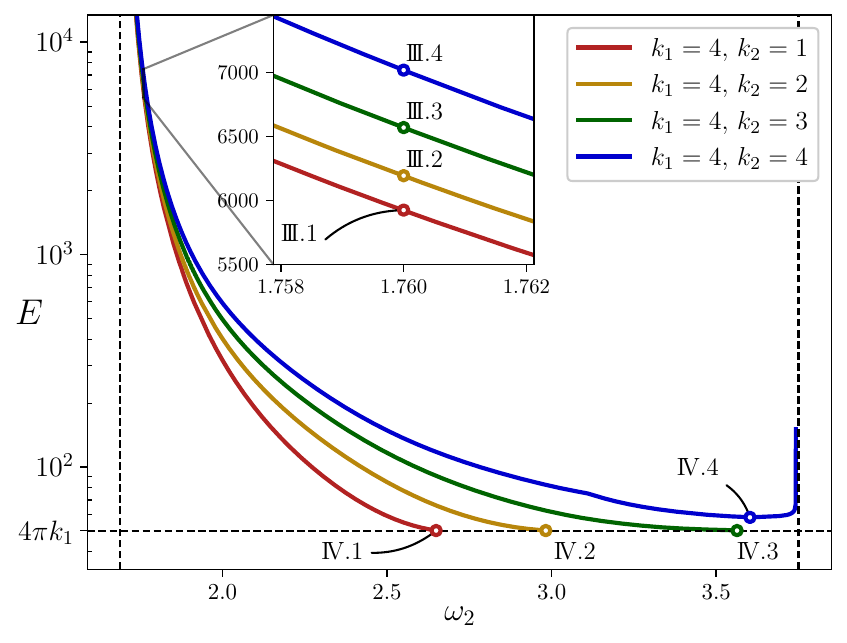}%
	\includegraphics[width=0.45\textwidth]{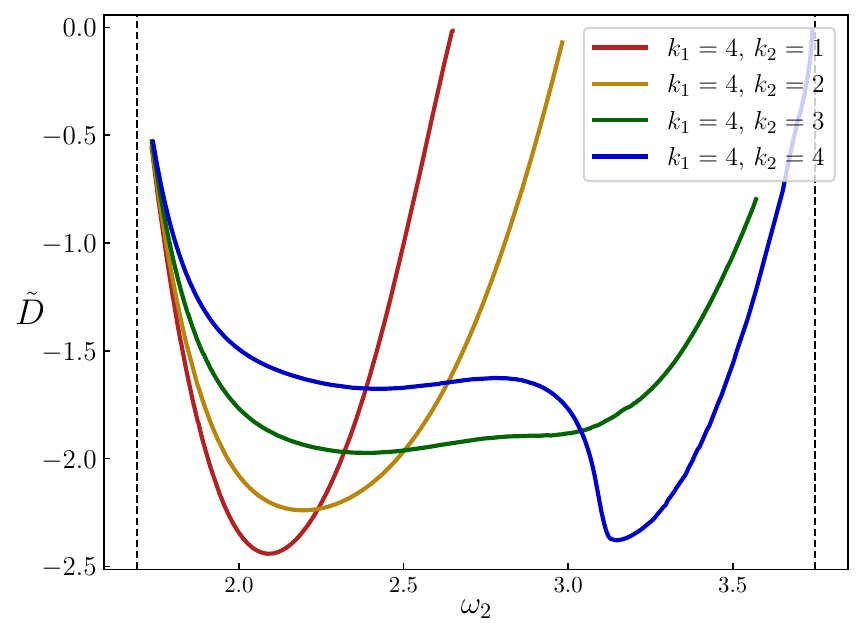}
	\caption{Total energy and normalized D-term of the $k_{1}=4$ $\mathbb{C}P^{2}$ solitons as functions of $\omega_{2}$ for fixed $\omega_{1}=1.00$. Dotted lines indicate the boundaries of allowed domain of values of the frequencies, cf Fig.~\ref{figs_Ew1w2_k1_1_k2_1}.}\label{figs_E_D_k1_4_k2_n}
\end{figure}

As a particular example of 
the $\mathbb{C}P^{2}$ solitons of higher degrees, we consider $Q_{\textup{top}}=4$ configurations with $k_1=4$ and set of values of non-topological integer $k_2=1,2,3,4$. 
Again, we fix the angular frequency $\omega_1=1$ and scan the allowed ranges of values of the second frequency $\omega_2$.
Generally, the pattern is similar to the situation described above with solitons of topological degree one. In Fig.~\ref{figs_E_D_k1_4_k2_n} we displayed the total energy of the $k_1=4$ solutions (left plot) and the corresponding normalized D-term (right plot) at $\omega_1=1$ vs
the angular frequency $\omega_2$  for a set of values of the second integer
$k_2$. The energy of configuration and the upper critical value of the frequency $\omega_2$  increases as the values of  $k_2$ increase, see Fig.~\ref{figs_E_D_k1_4_k2_n}, left plot. As $k_1 > k_2$,
both the energy and the angular momentum of the
configuration increase monotonically as the second frequency $\omega_2$ decreases. For the $k_1=k_2=4$ soliton, however, they diverge as $\omega_2$ approaches the threshold, then the Vakhitov-Kolokolov inequality does not hold. 

\begin{figure}[!htb]
	\centering
	\includegraphics[width=0.45\textwidth]{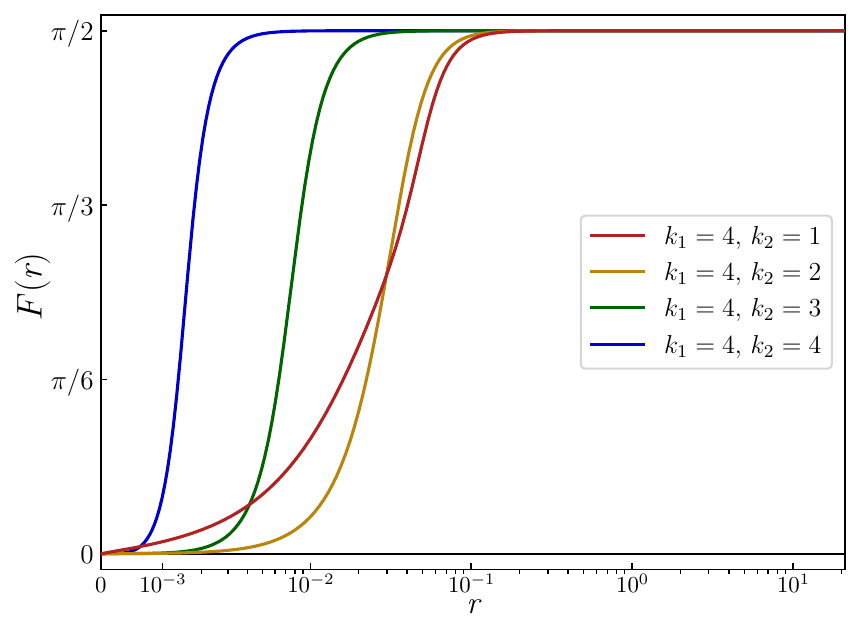}%
	\includegraphics[width=0.45\textwidth]{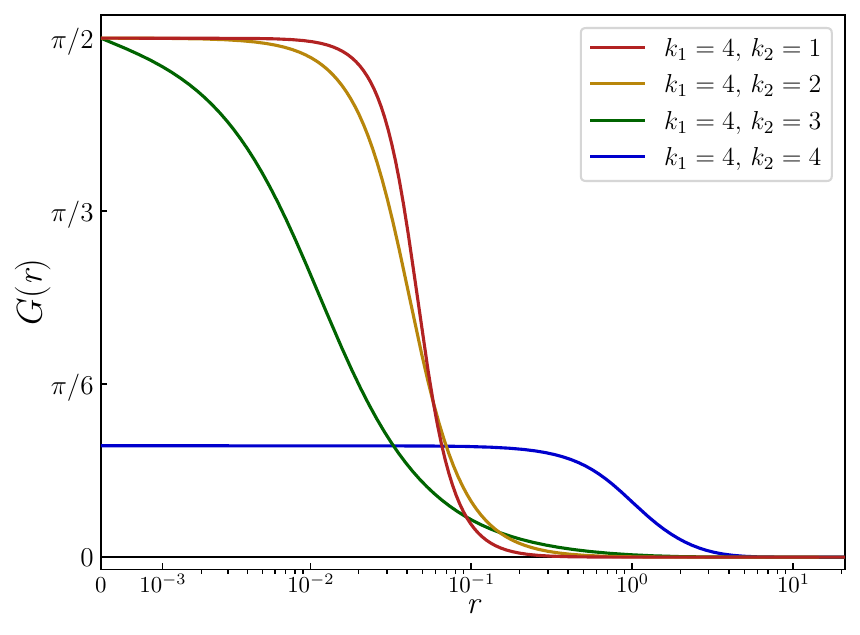}
	\caption{Radial profile functions of the $k_1 = 4$ $\mathbb{C}P^{2}$ solitons at $\omega_1 = 1$ for the set of allowed values of  $k_2=1-4$, labeled
as $\textup{I\!V}.k_{2}$ in Fig.~\ref{figs_E_D_k1_4_k2_n}, left plot (thick-wall limit). The corresponding values of $\omega_2$ are 
$2.649, 2.982, 3.563$ and $3.602$, respectively.
}\label{FG_4n_thick-wall_profiles}
\end{figure}

Similar to the case of solutions with topological charge one, considered above, there are thick and thin wall limits for solitons with topological charge four, see Figs.~\ref{FG_4n_thick-wall_profiles},\ref{FG_4n_thin-wall_profiles}. For the fixed value of the frequency $\omega_1=1$ the thick-wall $k_1=4$ configurations correspond to the values of the second frequency $\omega_2$ approaching the maximal critical values.

In Fig.~\ref{FG_4n_thick-wall_profiles}  we present radial profiles of the field functions $F(r)$ and $G(r)$ of the particular $k_1=4$ solutions in the thick-wall limit at $\omega_1=1$, labeled as $\textup{I\!V}.1-\textup{I\!V}.4$ in Fig.~\ref{figs_E_D_k1_4_k2_n}, left plot. The corresponding values of the second frequency for $k_2=1-4$ are 
$\omega_{2}= 2.649, 2.982, 3.563$ and $3.602$, respectively.

Note the different behavior of the profile function $ G(r)$ at the origin, the boundary conditions \re{BC1_FG} and  \re{BC1_FG_k1} are different for the $Q_{\textup{top}}=4$ $\mathbb{C}P^2$ solitons with $k_1=4>k_2$ and $k_1=k_2=4$. 

\begin{figure}[!htb]
	\centering
\includegraphics[width=0.45\textwidth]{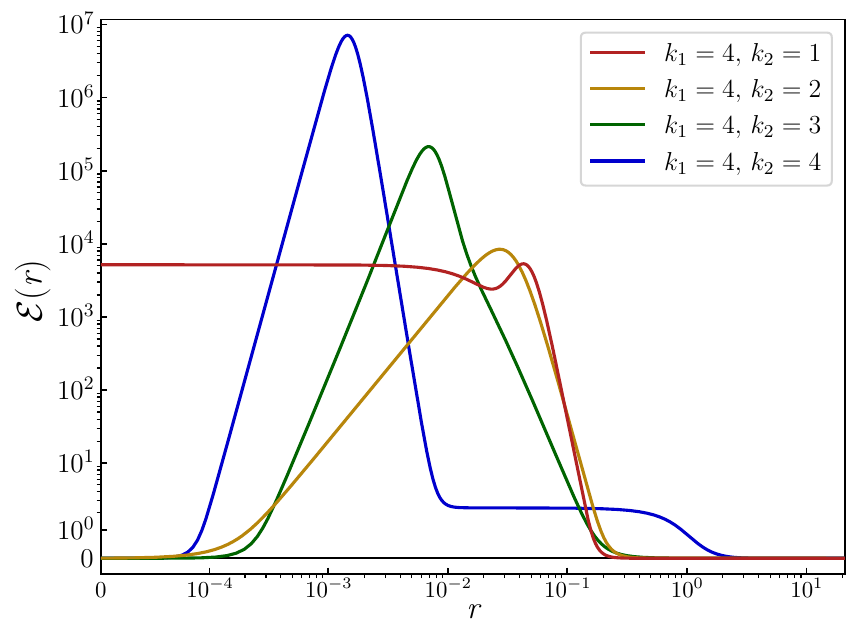}
\includegraphics[width=0.45\textwidth]{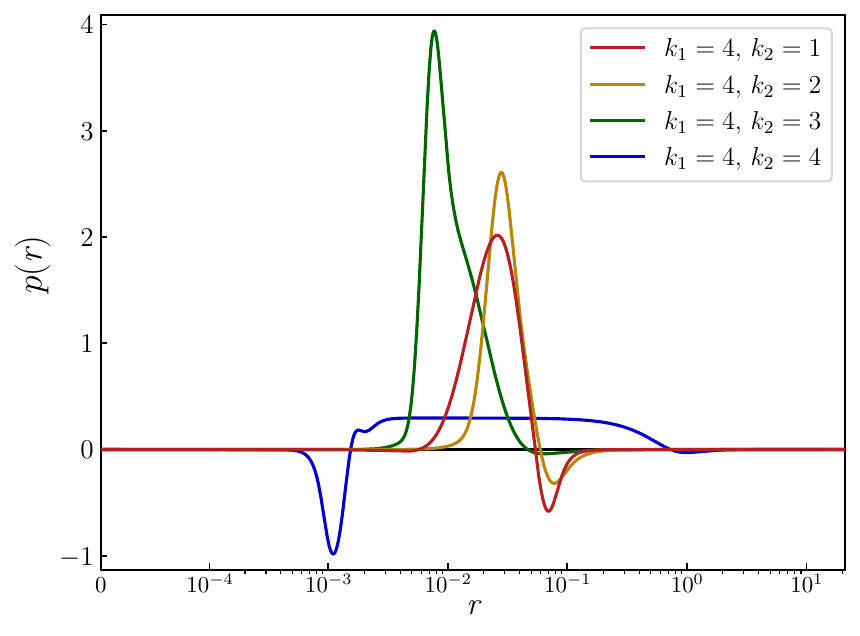}\\
	\includegraphics[width=0.45\textwidth]{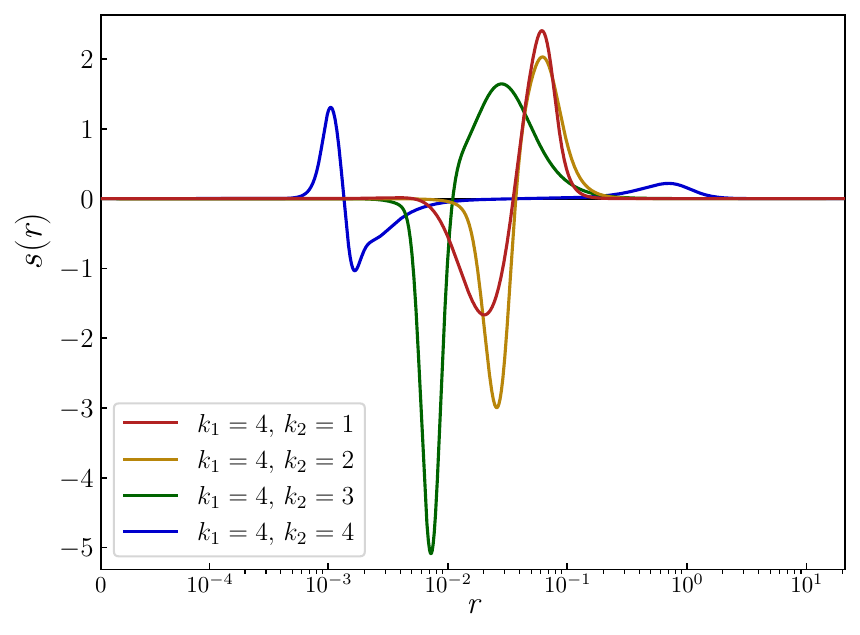} \includegraphics[width=0.45\textwidth]{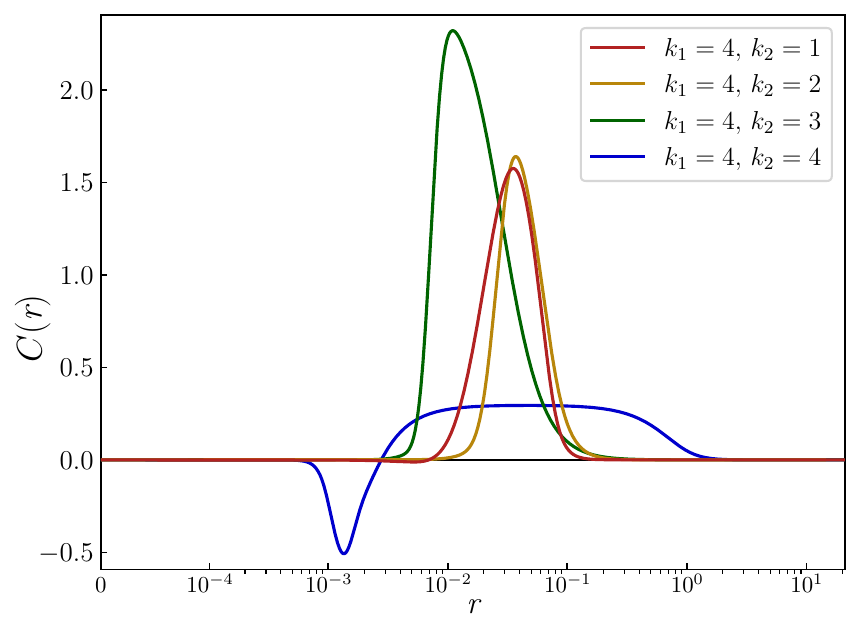}

	\caption{Radial profiles of the energy density distribution $\mathcal{E}(r)$ (upper left plot), the pressure function $p(r)$ (upper right plot), and shear force $s(r)$ (bottom left plot),  and the radial force component $C(r)$ (bottom right plot)  of the particular $k_{1}=4$ solutions at $\omega_1 =1$ for the set of allowed values of  $k_2=1-4$, labeled
as $\textup{I\!V}.k_{2}$ in Fig.~\ref{figs_E_D_k1_4_k2_n}, left plot (thick-wall limit). The corresponding values of $\omega_2$ are 
$2.649, 2.982, 3.563$ and $3.602$, respectively.
} 
\label{fig10}
\end{figure}

\begin{figure}[!htb]
	\centering
	\includegraphics[width=0.45\textwidth]{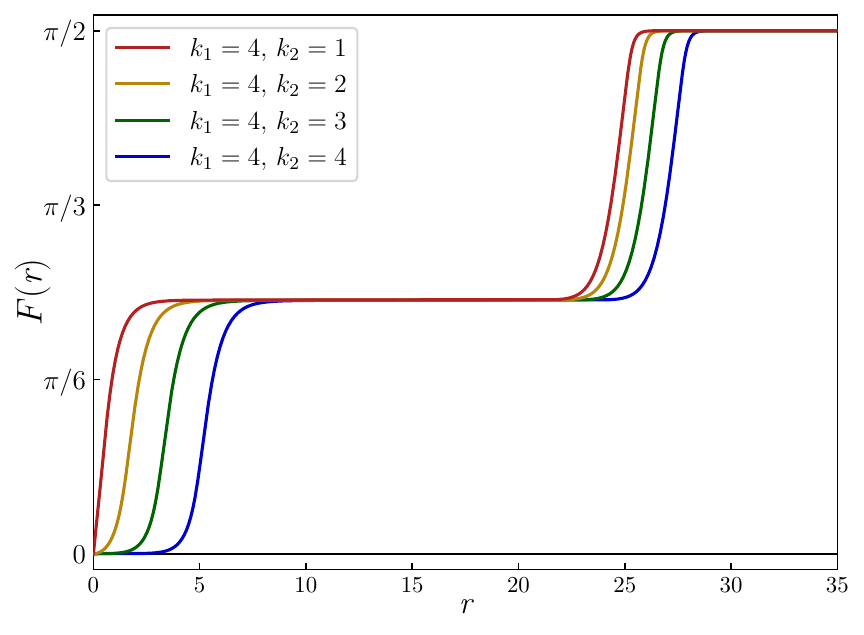}%
	\includegraphics[width=0.45\textwidth]{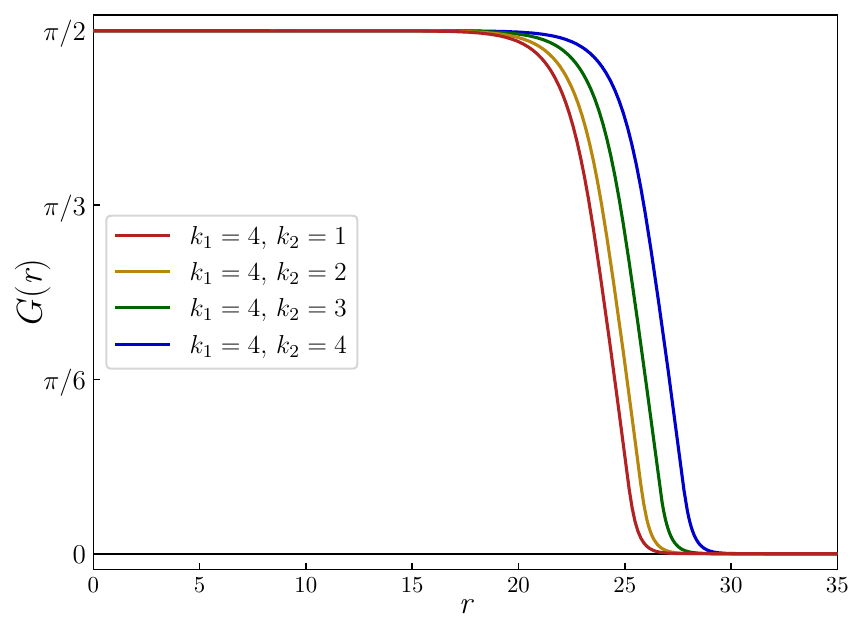}
	\caption{Radial profile functions of the $k_1 = 4$ $\mathbb{C}P^{2}$ solitons at $\omega_1 = 1$ for the set of allowed values of  $k_2=1-4$, labeled
as $\textup{I\!I\!I}.k_{2}$ in Fig.~\ref{figs_E_D_k1_4_k2_n}, left plot (thin-wall limit). The corresponding value of $\omega_2$ for all configurations is
$\omega_2=1.760$.    
}\label{FG_4n_thin-wall_profiles}
\end{figure}

The distributions of the energy density $\mathcal{E}(r)$ is different for configurations with $k_2>1$ and $k_2=1$, see Fig.\ref{fig10}, upper left plot. In the former case the function
of the radial energy distribution tends to zero both at the center of the configuration and at the spatial boundary having a pronounced maximum at some distance from the center, and an extended tail. For the $k_2=1$ configuration the energy distribution has a finite value at the origin, a small maximum at some distance from the center corresponds to the contribution of the surface energy. 

Also the radial distributions of the pressure $p(r)$ are different for the   
$Q_{\textup{top}}=4$ $\mathbb{C}P^2$ solitons with 
$k_2<4$ and $k_2=4$, see Fig.\ref{fig10}, upper right plot. For solutions with $k_2=1,2,3$, the von Laue criterion is not violated, but for solutions with $k_2=4$, the pressure function obviously has two zeros. The shear force $s(r)$ of configurations with $k_2<4$ becomes negative inside the core of the solitons, it is positive in the outer region and approaches zero at the spatial boundary from above, see \ref{fig10}, bottom left plot. The criteria function
$C(r)$ is positive for $k_2=1,2,3$ solitons, however, it becomes negative for $k_2=4$ configurations. 
Hence, we may conclude the $k_1=4$, $k_2=1,2,3$ configurations are stable in the thick-wall limit, while for the 
$k_1=4$, $k_2=4$ solitons the stability  criteria becomes violated although the D-term always remains negative, see  
Fig.~\ref{figs_E_D_k1_4_k2_n}, right plot.

\begin{figure}[!htb]
	\centering
	\includegraphics[width=0.45\textwidth]{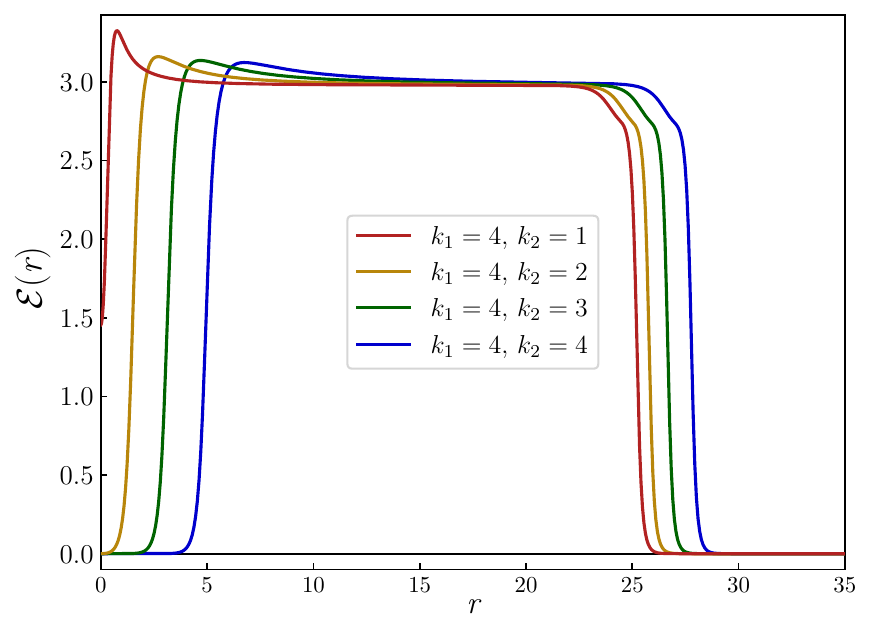}%
	\includegraphics[width=0.45\textwidth]{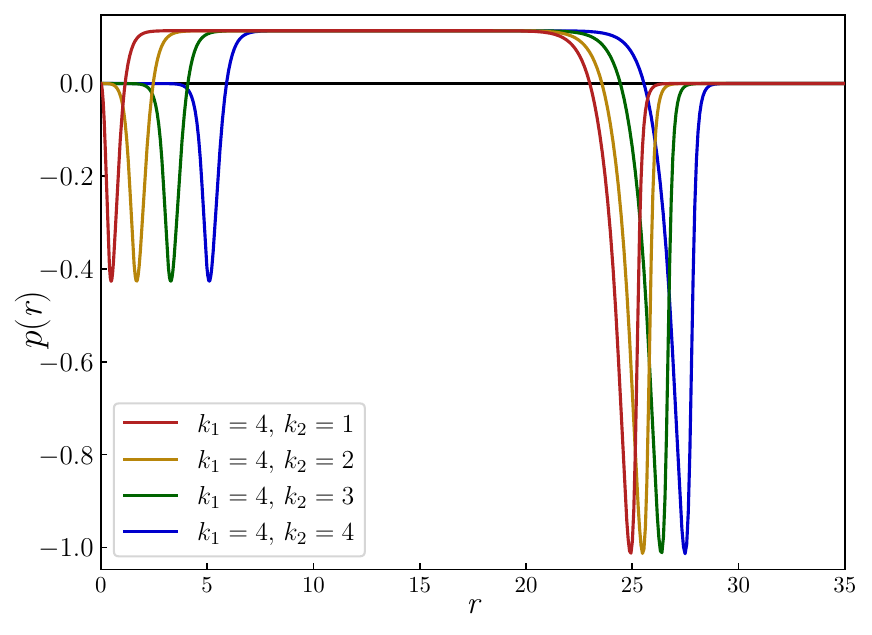}\\
	\includegraphics[width=0.45\textwidth]{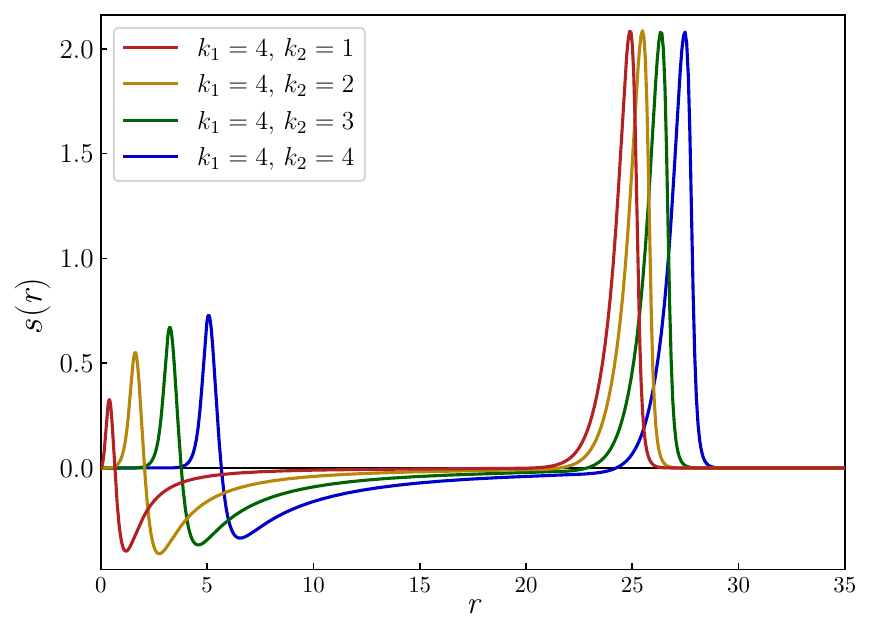}%
	\includegraphics[width=0.45\textwidth]{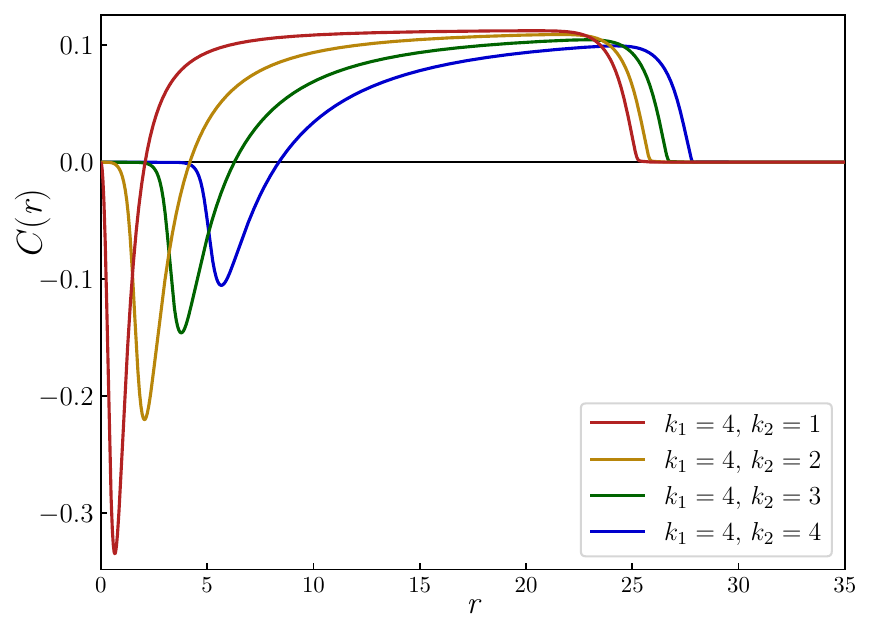}
	\caption{
    Radial profiles of the energy density distribution $\mathcal{E}(r)$ (upper left plot), the pressure function $p(r)$ (upper right plot), and shear force $s(r)$ (bottom left plot),  and the radial force component $C(r)$ (bottom right plot)  of the particular $k_{1}=4$ solutions at $\omega_1 =1$ for the set of allowed values of  $k_2=1-4$, labeled
as $\textup{I\!I\!I}.k_{2}$ in Fig.~\ref{figs_E_D_k1_4_k2_n}, left plot (thin-wall limit). The corresponding value of $\omega_2$ for all configurations is
$\omega_2=1.760$.    
}\label{EspC_4n_thin-wall_profiles}
\end{figure}

In the thin-wall limit the pattern becomes more simple. 
We demonstrate this
behavior of the $k_{1}=4$ $\mathbb{C}P^{2}$ solitons in Figs.~\ref{FG_4n_thin-wall_profiles} and \ref{EspC_4n_thin-wall_profiles}, where
we exhibit the radial profile functions and the distributions of the energy, pressure and shear forces of the configurations at $\omega_1=1$ in the thin-wall limit. For $\omega_2=1.760$ 
this configuration is clearly divided into two “half-solitons”, see Fig.~\ref{FG_4n_thin-wall_profiles}. The "topological" function $F(r)$ interpolates between the vacuum $F(0)=0$ at the origin and the value $\pi/4$. In this region the second profile function $G(r)$ approaches the value $\pi/2$, it corresponds to $n^3=\frac12, ~ n^8=-\frac{1}{2\sqrt{3}}$, i.e. the "anti-vacuum" with $W(n^3,n^8) = \frac{2}{3}$. In the far field region the components $F,G$ rapidly approach the vacuum on a compact domain. 
The size of the soliton core increases with the value of $k_2$. 

\begin{figure}[!htb]
	\centering
	\includegraphics[width=0.45\textwidth]{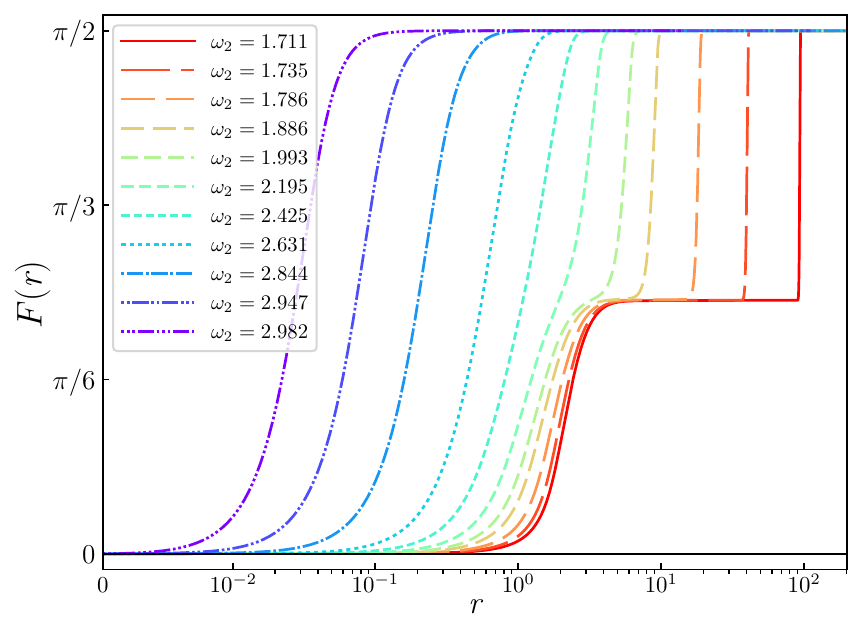}%
	\includegraphics[width=0.45\textwidth]{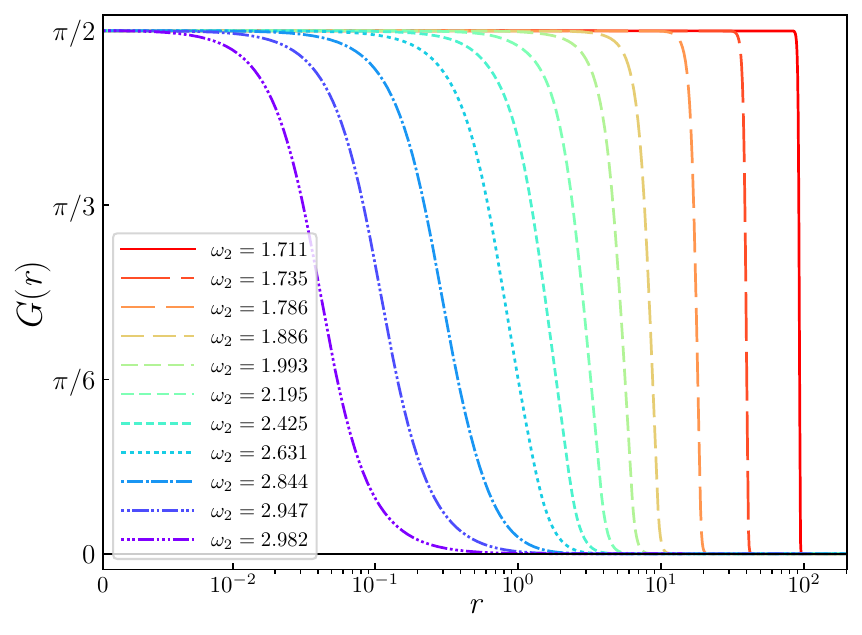}
	\caption{Radial profile functions $F(r), G(r)$ of the particular $k_1=4, ~k_2=2$ solitons at $\omega_1=1$ for a set of values of $\omega_{2}$.}\label{FG_continuum}
\end{figure}

\begin{figure}[!htb]
	\centering
	\includegraphics[width=0.45\textwidth]{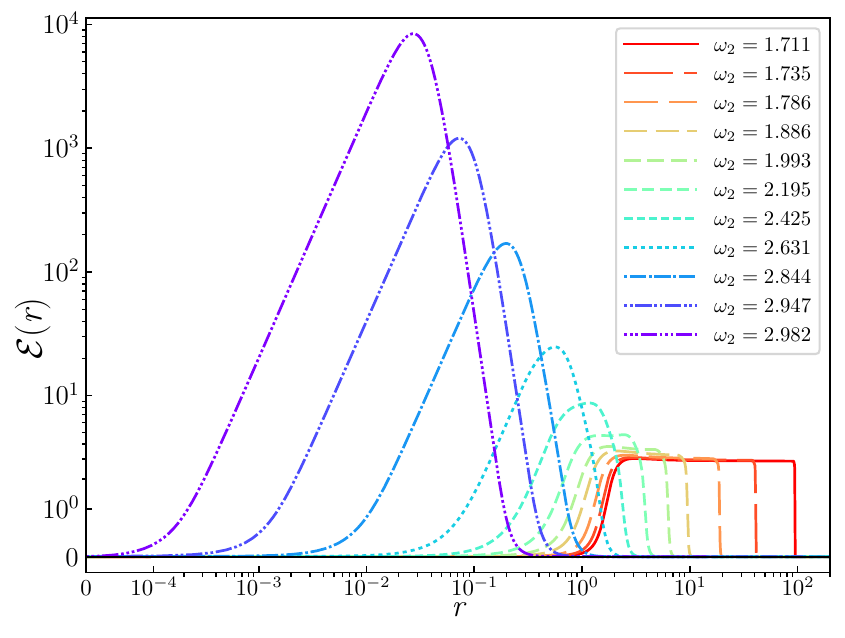}%
	\includegraphics[width=0.45\textwidth]{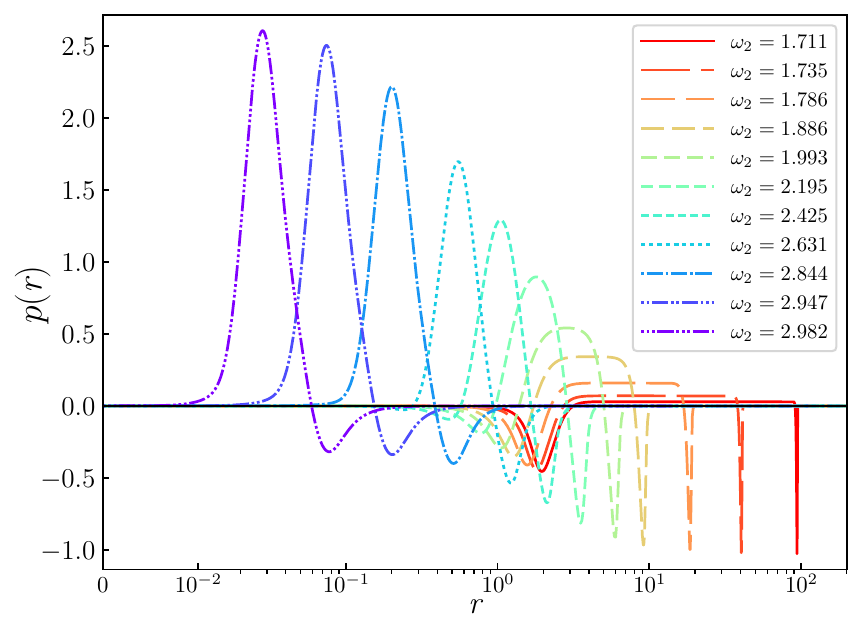}\\
	\includegraphics[width=0.45\textwidth]{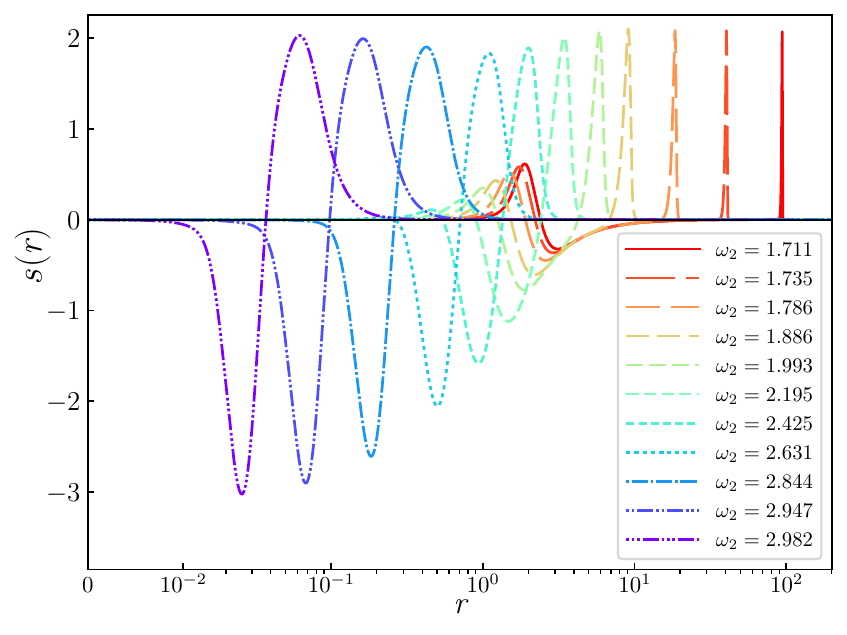}%
	\includegraphics[width=0.45\textwidth]{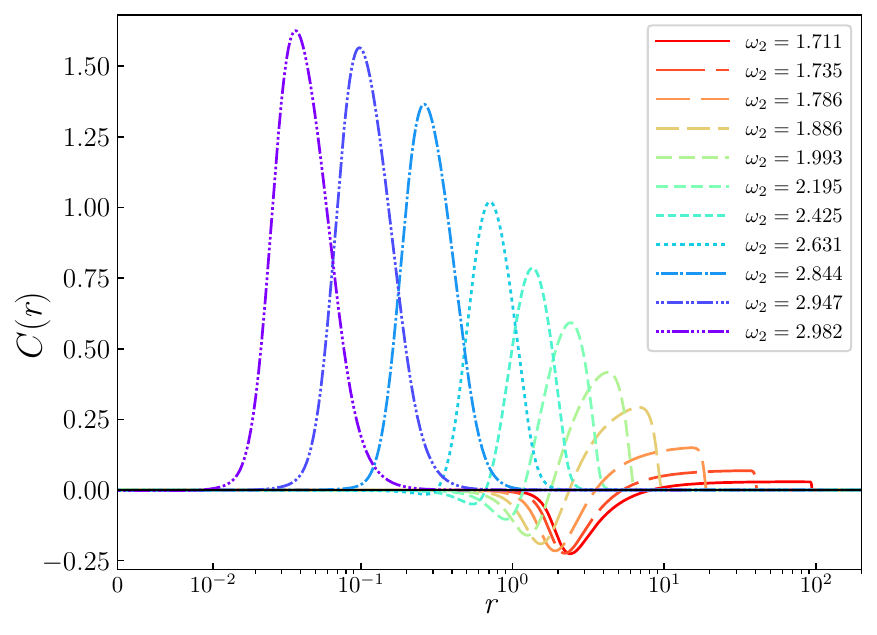}
	\caption{Radial profiles of pressure, shear stress, energy density and radial force component \eqref{C-criterion} of the particular $k_1=4, ~k_2=2$ solitons at $\omega_1=1$ for a set of values of $\omega_{2}$}\label{EspC_continuum}
\end{figure}

In Fig.~\ref{EspC_4n_thin-wall_profiles}, we present 
radial distributions of the total energy density $\mathcal{E}(r)$  and the functions $p(r)$, $s(r)$ and the normal force $C(r)$  of
the particular thin-wall  solutions, labeled as $\textup{I\!I\!I}.k_{2}$ in Fig.~\ref{figs_E_D_k1_4_k2_n}, left plot. Note that, similar to the case of $k_2=1$ soliton, the energy density distribution of the $k_1=1$ configuration has a finite value at the origin, when $k_2=2,3,4$ it tends to zero both at the origin and in the spatial asymptotic. For all solutions, the pressure and shear force functions $p(r)$, $s(r)$ clearly have two radial nodes, which correspond to the position of the two "half-solitons". The stability criterion function $C(r)$ is negative in the region close to the origin, it becomes positive for the second “half-soliton”, approaching zero on the spatial boundary from above. Therefore, we can conclude that these solutions are stress unstable.

Finally, we will consider the evolution patterns of $Q_{\textup{top}}=4$ $\mathbb{C}P^2$ solitons between the limiting cases of thin and thick walls. In Fig.~\ref{FG_continuum}  we displayed the profile functions of the 
$k_1=4, ~k_2=2$ solitons at $\omega_1=1$ for a set of values of $\omega_{2}$. 
Corresponding radial distributions of the total energy density $\mathcal{E}(r)$  and the functions $p(r)$, $s(r)$ and the normal force $C(r)$ are displayed in Fig.~\ref{EspC_continuum}. A configuration that is initially stable in the thick-wall limit becomes unstable with decreasing frequency $\omega_2$ and the formation of "half-solitons" as the thin-wall limit is approached.

\section{Conclusions and outlook}

In this paper we have considered the problem of classical stability of the isospinning 
$\mathbb{C}P^2$ 
topological solitons previously constructed in  
\cite{Amari:2024pnw}. 

Our approach to the study of soliton stability is based
on a treatment of the  $\mathbb{C}P^2$
topological soliton as an elastic medium and study of the corresponding matrix elements of the energy-momentum tensor, which contain information on the spatial distribution of internal forces acting within the configuration \cite{Polyakov:2002yz,Polyakov:2018zvc,Mai:2012yc,Mai:2012cx,Loiko:2022noq,Panteleeva:2023aiz,Farakos:2025byy,Mikhaliuk:2026tdt}.
We have derived expressions for the distribution of pressure and shear forces acting in the interior of a localized configuration.
Our results indicate that for the isospinning $\mathbb{C}P^2$ soliton the von Laue stability condition is not always satisfied. Considering the local stability criteria, we found  that these solitons become classically unstable for some range of values
of the parameters of the system. On the other hand, the corresponding $\mathbb{C}P^2$ D-term, which characterizes the distribution of internal forces \cite{Polyakov:2002yz,Polyakov:2018zvc,Perevalova:2016dln}, is always negative. By considering the branches of solutions that arise when the rotational frequencies decrease below the upper limiting values, we analyze the transition from the thin-wall and thick-wall limits.
An interesting observation is that in the thin-wall limit the stability criteria are always violated and the  configuration splits into two “half-solitons”. 
This may indicate that the rotationally invariant parameterization  \re{Z} is not fully consistent, the  profile functions of the soliton may be dependent on the angular variable.

Looking ahead, several promising avenues for further
study emerge. A natural next step will be to analyze the internal mechanical properties of non-rotating solitonic solutions of the $\mathbb{C}P^2$ model with stabilizing terms, like Skyrme term or with Dzyaloshinskii–Moriya interaction \cite{Akagi:2021dpk,Amari:2022boe}. Also non-topological iso-rotating solitons may exist in the $\mathbb{C}P^2$ model model with various potentials  \cite{Klimas:2017eft,Klimas:2021eue}.

Finally, we note that various interesting features of the topological and non-topological solutions of the $\mathbb{C}P^2$ model remain to be studied, in particular, there should be numerous
radially and angularly excited solutions. We hope to return elsewhere with a discussion of some of these interesting problems.
 
\section*{Acknowledgments}
We thank Yuki Amari, Luiz Ferreira, Jutta Kunz, Muneto Nitta, Emin Nugaev and Eugen Radu for useful discussions. Y.S. thanks Carlos Herdeiro and the Gr@v group at the University of Aveiro for their warm hospitality during the completion of this work and
gratefully acknowledges the support from the 
Portuguese Foundation for Science and
Technology (FCT—Fundacão para a Ciência e a
Tecnologia) 
mobility programme, grant RE-C06-i06.M02.   \newpage


 \begin{small}
 
 \end{small}

 \end{document}